\documentclass[12pt,a4paper]{article}
\usepackage{a4wide,cite}
\usepackage{amsmath,amssymb}
\usepackage{latexsym,graphicx,epsfig}

\newcommand{\half}{{\textstyle\frac{1}{2}}}

\newcommand{\threefourth}{{\textstyle\frac{3}{4}}}
\newcommand{\fourth}{{\textstyle\frac{1}{4}}}

\def\lsim{\mathrel{\rlap{\raise 2.5pt \hbox{$<$}}\lower 2.5pt\hbox{$\sim$}}}
\def\gsim{\mathrel{\rlap{\raise 2.5pt \hbox{$>$}}\lower 2.5pt\hbox{$\sim$}}}

\renewcommand{\Re}{{\rm Re\thinspace}}
\renewcommand{\Im}{{\rm Im\thinspace}}

\allowdisplaybreaks %!!!!
\begin{document}
\renewcommand{\thefootnote}{\fnsymbol{footnote}}
\newpage\normalsize
    \pagestyle{plain}
    \setlength{\baselineskip}{4ex}\par
    \setcounter{footnote}{0}
    \renewcommand{\thefootnote}{\arabic{footnote}}
\newcommand{\preprint}[1]{%
\begin{flushright}
\setlength{\baselineskip}{3ex} #1
\end{flushright}}
\renewcommand{\title}[1]{%
\begin{center}
    \LARGE #1
\end{center}\par}
\renewcommand{\author}[1]{%
\vspace{2ex}
{\Large
\begin{center}
    \setlength{\baselineskip}{3ex} #1 \par
\end{center}}}
\renewcommand{\thanks}[1]{\footnote{#1}}
\begin{flushright}
    \today
\end{flushright}
\vskip 0.5cm

\begin{center}
{\bf \Large
CP violation in top quark production at the LHC
and Two-Higgs-Doublet Models}
\end{center}
\vspace{1cm}
\begin{center}
Wafaa Khater  {\rm and}
Per Osland
\end{center}
%-----------------------------------
%   Address
%-----------------------------------
\vspace{1cm}
\begin{center}
Department of Physics,
University of Bergen,
    Allegt.\ 55, N-5007 Bergen, Norway
\end{center}
\vspace{1cm}
%%%%%%%%%%%%%%%%%%%%%%%%%%%%%%%%%%%%%%%%%%%%%%%%
\begin{abstract}
We discuss $CP$ violation in top-antitop production at the LHC,
induced by gluon fusion and final-state Higgs exchange.
Results by Bernreuther and Brandenburg are confirmed and further reduced.
The lepton energy asymmetry is studied in detail in explicit
Two-Higgs-Doublet Models with near-maximal mixing in the neutral Higgs sector.
Unless there is {\it only one light Higgs particle},
and unless (in ``Model II'') $\tan\beta\lsim1$, 
the $CP$-violating effects are very small,
possibly too small to be seen at the LHC.
\end{abstract}
%%%%%%%%%%%%%%%%%%%%%%%%%%%%%%%%%%%%%%%%%%%%%%%%%%%%%%%%%%%%%%%%%%%%%%%%
\section{Introduction}
\setcounter{equation}{0}
%%%%%%%%%%%%%%%%%%%%%%%%%%%%%%%%%%%%%%%%%%%%%%%%%%%%%%%%%%%%%%%%%%%%%%%%
One of the most promising ways in which one can search for new physics at the
LHC, is in $CP$ violation in connection with $t\bar t$ production.  It is
generally believed that the top quark, since it is very heavy, might be more
susceptible to new physics \cite{Schmidt:1992et,Atwood:1992vj}. 
In the particular case of Higgs-mediated
interactions, this is naturally the case, since the Higgs coupling to
the top quark is proportional to its mass.  This process
\begin{equation} \label{Eq:pp-to-ttbar}
pp \to t \bar t +X
\end{equation}
has been explored in considerable detail by Bernreuther and Brandenburg
\cite{Bernreuther:1994hq,Bernreuther:1998qv} who identified the different
kinematical structures appearing in the $CP$-violating part of the
interaction, and evaluated them in a generic Two-Higgs-Doublet Model
\cite{Lee:iz}.

Here, we review (and confirm) these calculations, and apply the results to a
particular version of the Two-Higgs-Doublet Model, in which the $CP$ violation
is minimal in structure \cite{Ginzburg:2001ss}.  This allows us to relate and
constrain the couplings and masses of the model. Such relations among
parameters are crucial in order to estimate the magnitude of possible signals.

If the Higgs states are not eigenstates under parity, then their couplings
to the fermions will violate $CP$. 
In particular, if the $Ht\bar t$ coupling (for a given Higgs particle)
is of the generic form
\begin{equation}
H t \bar t: \qquad 
[a+i\gamma_5\tilde a], 
\end{equation}
then the $CP$-violating part of the cross section for the process
(\ref{Eq:pp-to-ttbar}), which depends on the $t$ and $\bar t$ spins, will be
proportional to the dimensionless model-dependent quantity
\begin{equation}  \label{Eq:gamma_CP-def}
\gamma_{CP}=-a \tilde a
\end{equation}
where $a$, $\tilde a$ are the reduced scalar and pseudoscalar Yukawa
couplings, respectively.

Here, we restrict ourselves to the subprocess $gg\to t\bar t$ since it is the
leading $t\bar t$ production mechanism at the CERN LHC.  The $q\bar q$ initial
states, which are important at the Tevatron \cite{Bernreuther:1994hq},
contribute at the LHC less than 10\% to the total cross section
\cite{QCD-book} and give only a numerically negligible contamination of the
$CP$-violating signal \cite{Schmidt:1992et,Bernreuther:1994hq}.

The MSSM provides an alternative, very interesting framework for $CP$
violation via the Higgs sector \cite{Babu:1997jr}.
Additional effects in the MSSM include $CP$-violating gluino exchange
\cite{Schmidt:1992kt,Zhou:1998wz}
as well as those due to phases in 
the bilinear and trilinear couplings, given by the so-called $\mu$ and
$A_t$ parameters. For an application to the process (\ref{Eq:pp-to-ttbar}),
see \cite{Zhou:1998wz}.
We here restrict ourselves to the 2HDM, this
is already a rich framework.

The paper is organized as follows.  After a review of notation and relevant
formulas in sect.~2, we give model-independent results in sect.~3, focusing on
basic observables discussed in \cite{Bernreuther:1994hq}.  In sect.~4, we
review the physical content of the 2HDM, and in sect.~5, we study the
magnitude of the $CP$ violation for two distinct neutral Higgs mass spectra:
two light and one heavy {\it vs.}\ one light and two heavy. Sect.~6 is devoted
to concluding remarks, and an appendix summarizes the basic one-loop results
for the $CP$-violating amplitudes.
%%%%%%%%%%%%%%%%%%%%%%%%%%%%%%%%%%%%%%%%%%%%%%%%%%%%%%%%%%%%%%%%%%%%%%%
\section{Preliminaries}
\setcounter{equation}{0}
%%%%%%%%%%%%%%%%%%%%%%%%%%%%%%%%%%%%%%%%%%%%%%%%%%%%%%%%%%%%%%%%%%%%%%%
A schematic representation of a generic one-loop diagram of the
process $gg\to t\bar{t}$ is
given in fig.~\ref{Fig:1}.\footnote{To produce the figures,
the {\sf AXODRAW} package \cite{Vermaseren:1994je} was used.}
%%%%%%%%%%%%%%%%%%%%%%%%%%%%%%%%%%%%%%%%%%%%%%%%%%%%%%%%%%%%%%%%%%%%%%%%
\begin{figure}[htb]
\refstepcounter{figure}
\label{Fig:1}
\addtocounter{figure}{-1}
\begin{center}
\setlength{\unitlength}{1cm}
\begin{picture}(5.0,4.0)
\put(-2,0){
\mbox{\epsfysize=4cm
\epsffile{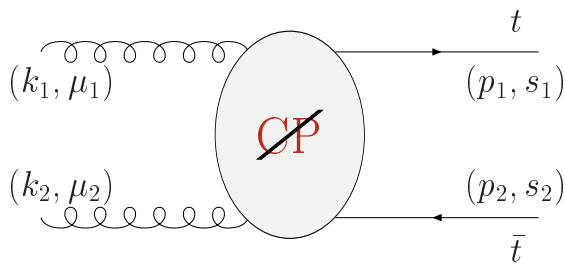}}}
\end{picture}
\vspace*{-4mm}
\caption{Kinematics of the underlying $g(k_1)+g(k_2)\to
t(p_1)+\bar{t}(p_2)$ reaction.}
\end{center}
\end{figure}
%%%%%%%%%%%%%%%%%%%%%%%%%%%%%%%%%%%%%%%%%%%%%%%%%%%%%%%%%%%%%%%%%%%%%%

%%%%%%%%%%%%%%%%%%%%%%%%%%%%%%%%%%%%%%%%%%%%%%%%%%%%%%%%%%%%%%%%%%%%%%%%
\subsection{Notation}
%%%%%%%%%%%%%%%%%%%%%%%%%%%%%%%%%%%%%%%%%%%%%%%%%%%%%%%%%%%%%%%%%%%%%%%%
The momentum and the spin four-vectors of the top (and antitop) quark 
are denoted by $p_1$ and $s_1$ (and $p_2$ and $s_2$), respectively,
with $m_t$ the mass,
whereas the momenta and the Lorentz indices of the initial gluons are
represented by $k_i$ and $\mu_i$ ($i=1,2$). We introduce the
linearly-independent set of momenta ($P_g$, $Q$, $P_t$): 
\begin{equation}
Q=k_1+k_2=p_1+p_2,\qquad
P_g=k_1-k_2,  \qquad
P_t=p_1-p_2,
\label{eq:art_varH}
\end{equation}
with $Q\cdot P_t=Q\cdot P_g=0$. The non-vanishing scalar products involving
the momenta $(P_g$, $Q$, $P_t)$ are given by $Q^2=-P_g^2=\hat s$,
$P_t^2=-\beta^2\hat{s}$ and $(P_g\cdot P_t)=-\beta z\hat{s}$ where $\hat s$ is
the gluon--gluon center of mass energy squared, $\beta=\sqrt{1-4m_t^2/\hat s}$
is the top quark velocity and $z=\cos\theta=(\hat{\bf P}_g\cdot\hat{\bf
P}_t)$, with $\theta$ the scattering angle in the gluon--gluon center of mass
frame. Working with the linearly-independent set of momenta ($P_g$, $Q$,
$P_t$) simplifies the kinematics, the four-point first- and
second-rank tensor loop integrals in particular.  
%%%%%%%%%%%%%%%%%%%%%%%%%%%%%%%%%%%%%%%%%%%%%%%%%%%%%%%%%%%%%%%%%%%%%%%%
\begin{figure}[htb]
\refstepcounter{figure}
\label{Fig:2}
\addtocounter{figure}{-1}
\begin{center}
\setlength{\unitlength}{1cm}
\begin{picture}(5.0,4.0)
\put(-3,0)
{\mbox{\epsfysize=4cm
\epsffile{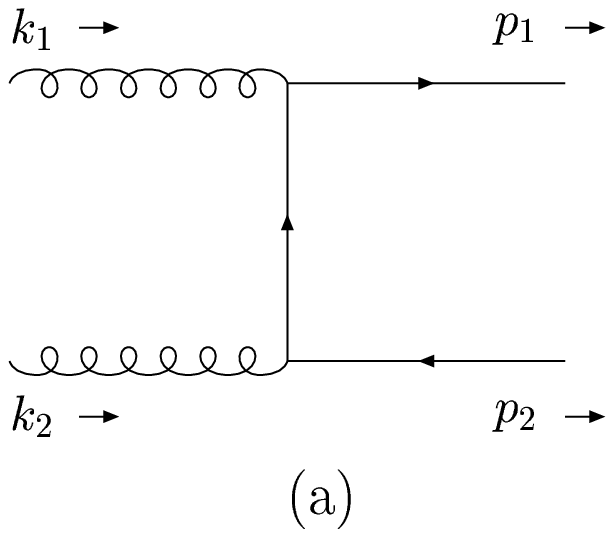}}
 \mbox{\epsfysize=4cm
\epsffile{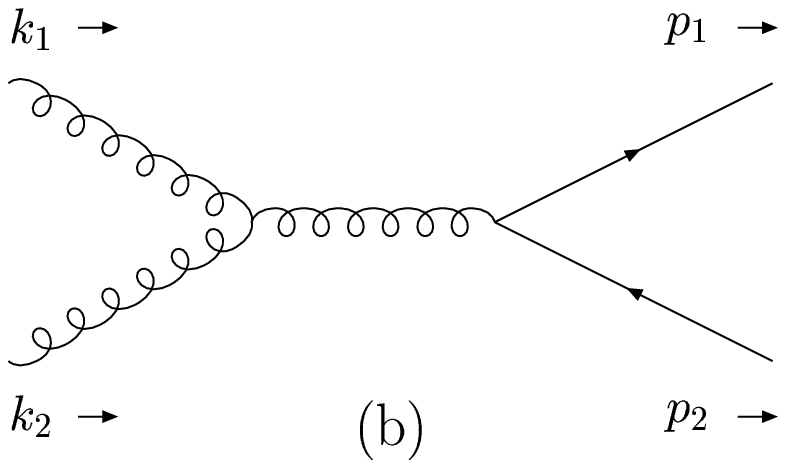}}}
\end{picture}
%\vspace*{-4mm}
\caption{Lowest-order QCD Feynman diagrams of the underlying $gg\to
t\bar{t}$ reaction, the crossed diagram of ($a$) is not shown.}
\end{center}
\end{figure}
%%%%%%%%%%%%%%%%%%%%%%%%%%%%%%%%%%%%%%%%%%%%%%%%%%%%%%%%%%%%%%%%%%%%%%

%%%%%%%%%%%%%%%%%%%%%%%%%%%%%%%%%%%%%%%%%%%%%%%%%%%%%%%%%%%%%%%%%%%%%%%%
\begin{figure}[htb]
\refstepcounter{figure}
\label{Fig:3}
\addtocounter{figure}{-1}
\begin{center}
\setlength{\unitlength}{1cm}
\begin{picture}(5.0,12.0)
\put(-3,8)
{\mbox{\epsfysize=4cm
\epsffile{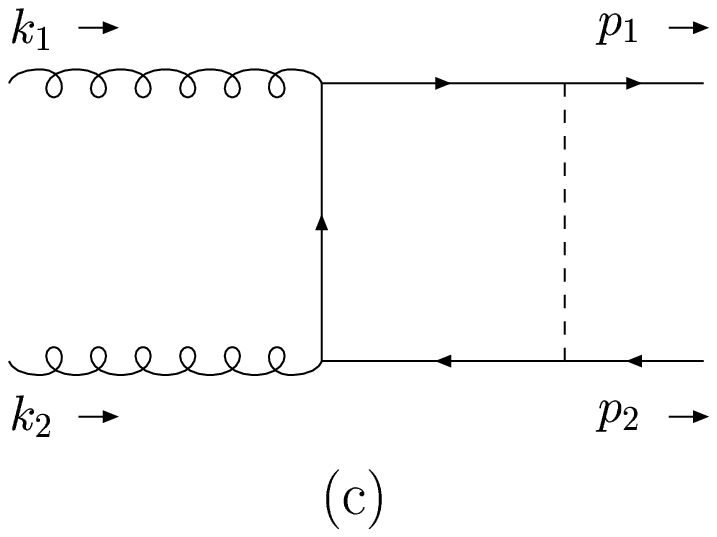}}
 \mbox{\epsfysize=4cm
\epsffile{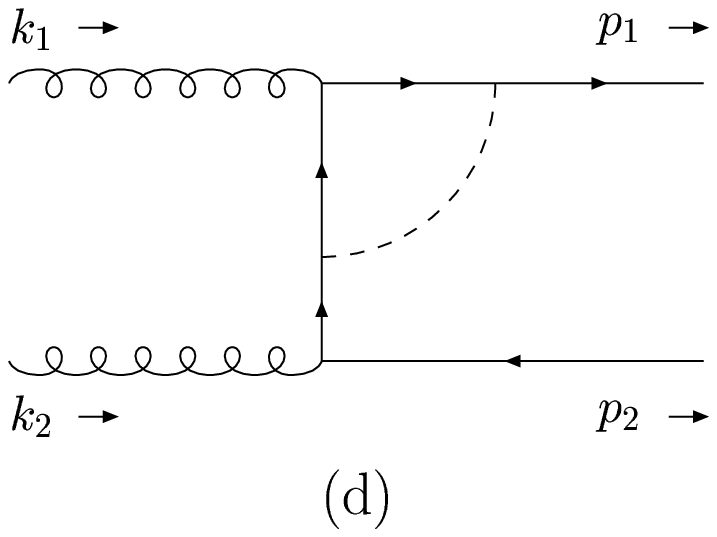}}}
\put(-3,4)
{\mbox{\epsfysize=4cm
\epsffile{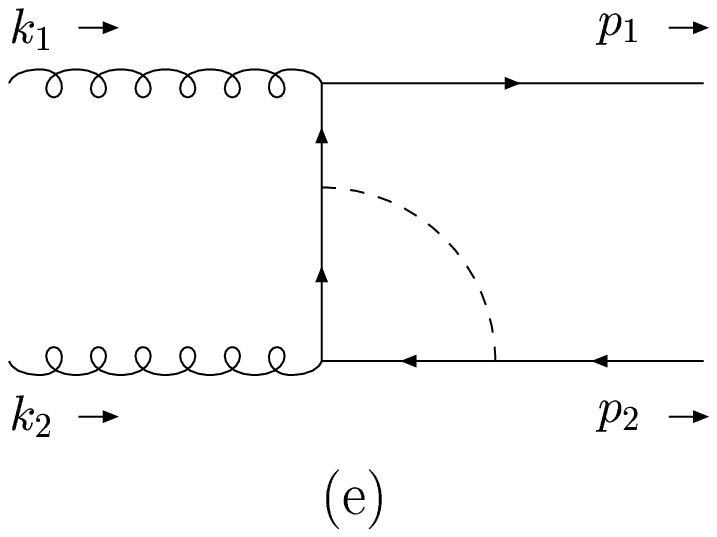}}
 \mbox{\epsfysize=4cm
\epsffile{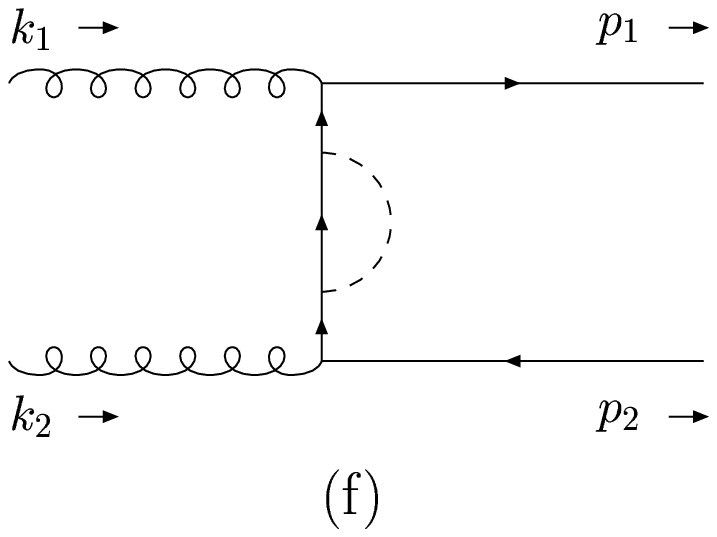}}}
\put(-3,0)
{\mbox{\epsfysize=4cm
\epsffile{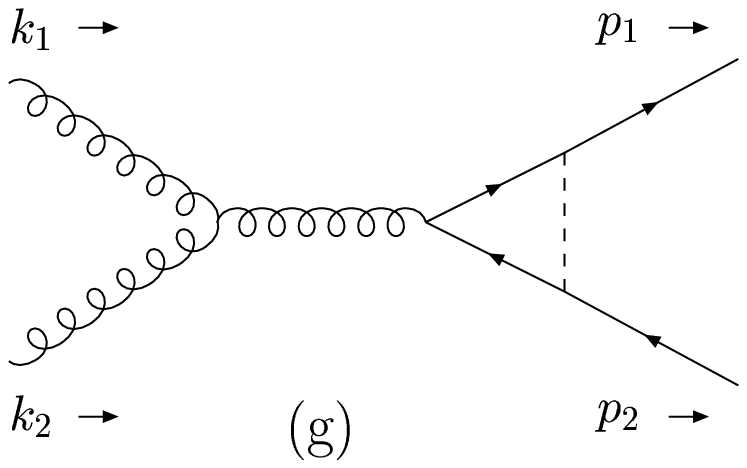}}
 \mbox{\epsfysize=4cm
\epsffile{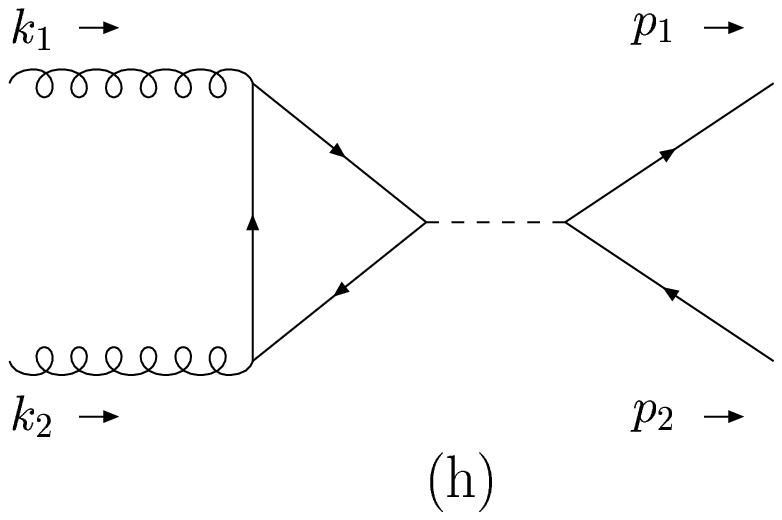}}}
\end{picture}
%\vspace*{-4mm}
\caption{Feynman diagrams of the underlying $gg\to t\bar{t}$ reaction with
neutral non-Standard-Model Higgs exchanges (dashed).  The crossed diagrams are
not shown (diagram $(g)$ has no crossed partner).}
\end{center}
\end{figure}
%%%%%%%%%%%%%%%%%%%%%%%%%%%%%%%%%%%%%%%%%%%%%%%%%%%%%%%%%%%%%%%%%%%%%%

Denoting by ${\cal M}_0$ the QCD amplitude corresponding to the diagrams of
fig.~\ref{Fig:2}, and by ${\cal M}_1$ the one corresponding to the one-loop
Higgs-exchange diagrams of fig.~\ref{Fig:3}, we obtain the cross section
\begin{equation}
\sigma\propto|{\cal M}_0+{\cal M}_1|^2.
\end{equation}
Spin-dependent $CP$-violating parts arise from terms linear in $\gamma_{CP}$,
(\ref{Eq:gamma_CP-def}).  Those originate from interference between ${\cal
M}_0$ and ${\cal M}_1$, and from $|{\cal M}_1|^2$.\footnote{When calculating
linear terms in $\gamma_{CP}$ in $|{\cal M}_1|^2$, we only considered the
contribution from diagram $h$ in fig.~\ref{Fig:3}, which is important for
$m_H$ above the $t\bar t$ threshold \cite{Bernreuther:1998qv}.} We write
\begin{equation}
\sigma=\sigma_{\rm even}+\sigma_{CP},
\end{equation}
where the non-interesting $CP$-even part of the cross section
$\sigma_{\rm even}$ results from $|{\cal M}_0|^2$ and from the terms
in $|{\cal M}_1|^2$ that are independent of $\gamma_{CP}$ or {\it even}
in this quantity.
The $CP$-violating part of the cross section can
be written in the most general Lorentz-invariant form as
\begin{align}  \label{eq:cross-sect}
\sigma_{CP}
&\propto\Phi_1[P_g\cdot s_1 - (1\leftrightarrow2)]
 +\Phi_2[P_t\cdot s_1 - (1\leftrightarrow2)]
 +\Phi_3[R\cdot s_1 - (1\leftrightarrow2)] \nonumber \\
&+\Psi_1[(R\cdot s_1)(P_g\cdot s_2) - (1\leftrightarrow2)]
 +\Psi_2[(R\cdot s_1)(P_t\cdot s_2) - (1\leftrightarrow2)] \nonumber \\
&+\Psi_3[(P_g\cdot s_1)(P_t\cdot s_2) - (1\leftrightarrow2)],
\end{align}
with the auxiliary pseudovector
\begin{equation}
R_\sigma=\epsilon_{\mu\nu\rho\sigma}P_g^\mu Q^\nu P_t^\rho.
\end{equation}
The coefficients $\Phi_1, \ldots \Phi_3$ and $\Psi_1, \ldots \Psi_3$, the
structure of which are determined by the actual Higgs exchange diagrams
(for details, see \cite{Bernreuther:1994hq,Wafaa}),
depend on $\hat s$, $z$ and the masses of the Higgs
boson and the top quark, $m_H$ and $m_t$.
%%%%%%%%%%%%%%%%%%%%%%%%%%%%%%%%%%%%%%%%%%%%%%%%%%%%%%%%%%%%%%%%%%%%%%%%
\subsection{The Bernreuther--Brandenburg decomposition}
%%%%%%%%%%%%%%%%%%%%%%%%%%%%%%%%%%%%%%%%%%%%%%%%%%%%%%%%%%%%%%%%%%%%%%%%
Bernreuther and Brandenburg describe the $CP$ violation in the process of
fig.~\ref{Fig:1} in terms of the $t\bar t$ production density matrix,
eq.~(2.8) in \cite{Bernreuther:1994hq} as:
\begin{equation} \label{eq:prod_mat}
{\cal R}_{CP}
=(b_{g1}^{CP}\hat k_i+b_{g2}^{CP}\hat p_i+b_{g3}^{CP}\hat n_i)
(\sigma^i\otimes\boldsymbol{1}-\boldsymbol{1}\otimes\sigma^i)
+\epsilon_{ijk}(c_{g1}\hat k_i+c_{g2}\hat p_i+c_{g3}\hat n_i)
\sigma^j\otimes\sigma^k,
\end{equation}
where (in our notation) ${\bf\hat k}={\bf\hat P}_g$, ${\bf\hat p}={\bf\hat
P}_t$ and ${\bf\hat n}$ is the unit vector in the direction of ${\bf n}={\bf
P}_g\times {\bf P}_t$ defined in the gluon--gluon center of mass
frame.\footnote{For the momenta of the partons involved, we use the notation
$g(k_1)+g(k_2)\to t(p_1)+\bar t(p_2)$, while in \cite{Bernreuther:1994hq} the
notation $g(p_1)+g(p_2)\to t(k_1)+\bar t(k_2)$ was used.} The $\sigma^i$ are
Pauli matrices with ${\bf s_1}= \half\boldsymbol{\sigma}\otimes\boldsymbol{1}$
(${\bf s_2}= \half\boldsymbol{1}\otimes\boldsymbol{\sigma}$) the spin operator
of the top (anti-top) defined in the top (anti-top) rest frame.

The relation of this expansion to our eq.~(\ref{eq:cross-sect}) is given by:
\begin{align}           \label{Eq:bg1-cg2}
b_{g1}^{CP}&=-\sqrt{\hat s}\,\Phi_1,
\nonumber \\
b_{g2}^{CP}
&=-\sqrt{\hat s}\left(\frac{\sqrt{\hat s}}{2m_t}-1\right)z\,\Phi_1
  -\frac{\hat s}{2m_t}\,\beta\,\Phi_2,
\nonumber \\
b_{g3}^{CP}&=\hat s\sqrt{\hat s}\,\beta\sqrt{1-z^2}\,\Phi_3,
\nonumber \\
c_{g1}&=\frac{\hat s^2\sqrt{\hat s}\,\beta}{2m_t}[z\Psi_1+\beta\,\Psi_2],
\nonumber \\
c_{g2}&=-\frac{\hat s^2\sqrt{\hat s}\,\beta z}{2m_t}[z\Psi_1+\beta\,\Psi_2]
-\hat s^2\beta (1-z^2)\Psi_1,
\nonumber \\
c_{g3}&=\frac{\hat s\sqrt{\hat s}}{2m_t}\beta\sqrt{1-z^2}\,\Psi_3.
\end{align}
The symmetry properties of these functions are given in
\cite{Bernreuther:1994hq}.

We confirm the results for these functions (\ref{Eq:bg1-cg2}) given in the
appendix of ref.~\cite{Bernreuther:1994hq}, up to some misprints mentioned in
\cite{Bernreuther:1998qv}.  Also, we have further reduced the results of the
box diagram to the basic four-point function together with three- and
two-point functions. This yields a more unique representation, and is also
convenient for the use of standard loop function libraries, like the
``LoopTools'' package \cite{Hahn:1998yk,vanOldenborgh:1989wn}.  Our results
are collected in Appendix~A.

%%%%%%%%%%%%%%%%%%%%%%%%%%%%%%%%%%%%%%%%%%%%%%%%%%%%%%%%%%%%%%%%%%%%%%%%
\section{Model-independent results} \label{sect:model-indep}
\setcounter{equation}{0}  \label{Sec:model-indep}
%%%%%%%%%%%%%%%%%%%%%%%%%%%%%%%%%%%%%%%%%%%%%%%%%%%%%%%%%%%%%%%%%%%%%%%%
Before invoking concrete models, we shall first update and review
some $CP$-violating quantities discussed in \cite{Bernreuther:1994hq},
both at the parton level and at the hadronic level.
%%%%%%%%%%%%%%%%%%%%%%%%%%%%%%%%%%%%%%%%%%%%%%%%%%%%%%%%%%%%%%%%%%%%%%%%
\subsection{CP violation at the parton level}
%%%%%%%%%%%%%%%%%%%%%%%%%%%%%%%%%%%%%%%%%%%%%%%%%%%%%%%%%%%%%%%%%%%%%%%%

We first consider the non-vanishing\footnote{The
corresponding quantities involving the gluon or beam direction, vanish
due to the symmetry properties under $z$.} parton-level $CP$-odd quantities
\cite{Bernreuther:1994hq}:
\begin{align}      \label{Eq:z*bg1+bg2}
\langle{\bf\hat p}\cdot(\boldsymbol{s}_1-\boldsymbol{s}_2)\rangle_g
&=\frac{4\int_{-1}^1 dz(zb_{g1}^{CP}+b_{g2}^{CP})}{4\int_{-1}^1 dz\,A^g}, \\
\langle{\bf\hat p}\cdot(\boldsymbol{s}_1\times\boldsymbol{s}_2)
\rangle_g
&=\frac{2\int_{-1}^1 dz(zc_{g1}^{CP}+c_{g2}^{CP})}{4\int_{-1}^1 dz\,A^g},
      \label{Eq:z*cg1+cg2}
\end{align}
where $A^g$ is the spin-independent part of the $CP$-even part of the
production density matrix given analytically in \cite{Bernreuther:1994hq}.
We note that these two spin--spin correlations are given by absorptive
and dispersive parts of the amplitudes (see Appendix~A).
Plots of these expectation values {\it vs.}\ the center of mass energy
$\sqrt{\hat s}$ are shown in fig.~\ref{Fig:bb-cc}.

%%%%%%%%%%%%%%%%%%%%%%%%%%%%%%%%%%%%%%%%%%%%%%%%%%%%%%%%%%%%%%%%%%%%%%%%
\begin{figure}[htb]
\refstepcounter{figure}
\label{Fig:bb-cc}
\addtocounter{figure}{-1}
\begin{center}
\setlength{\unitlength}{1cm}
\begin{picture}(10.0,7.0)
\put(-3.5,0)
{\mbox{\epsfysize=7.5cm
\epsffile{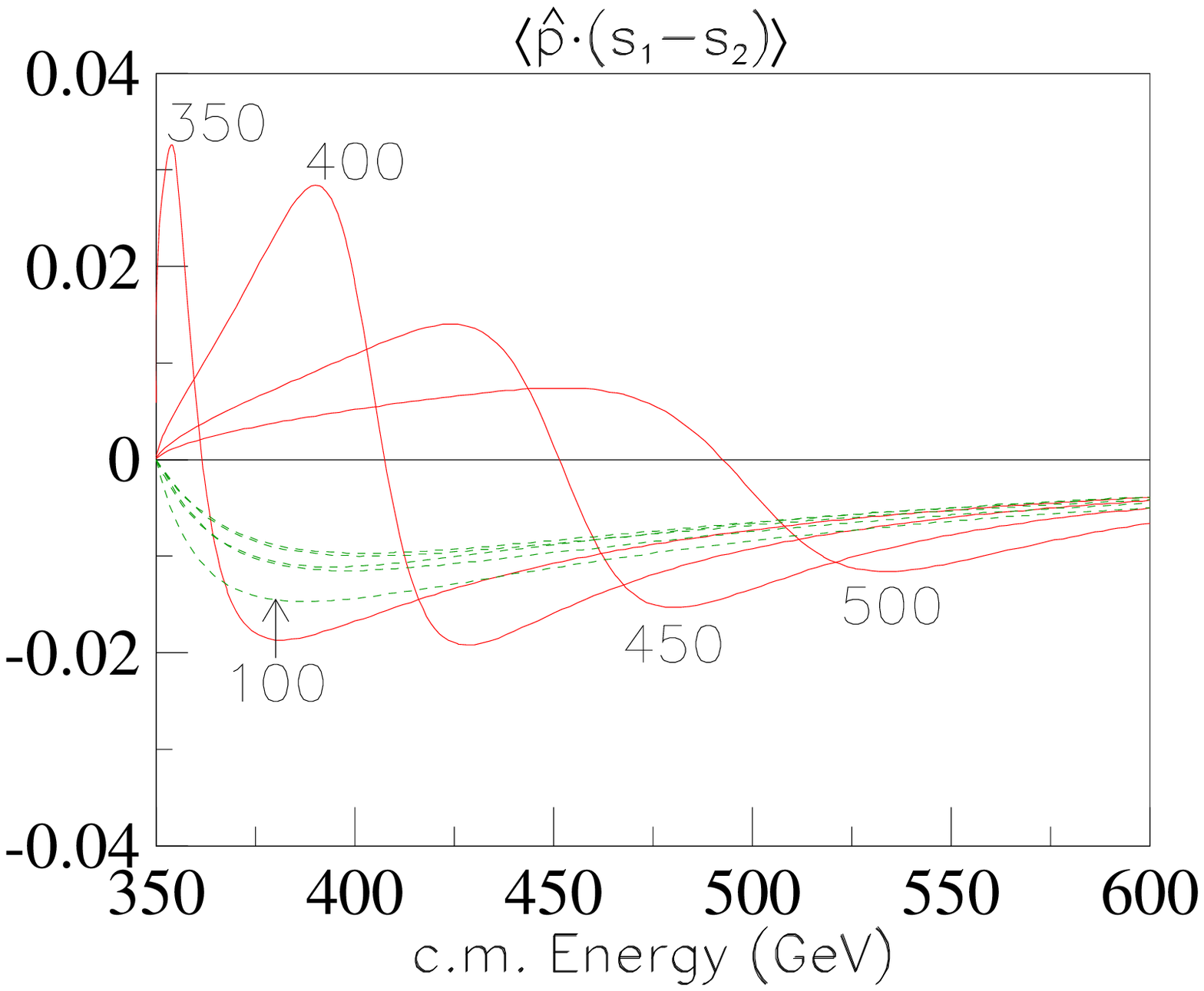}}
 \mbox{\epsfysize=7.5cm
\epsffile{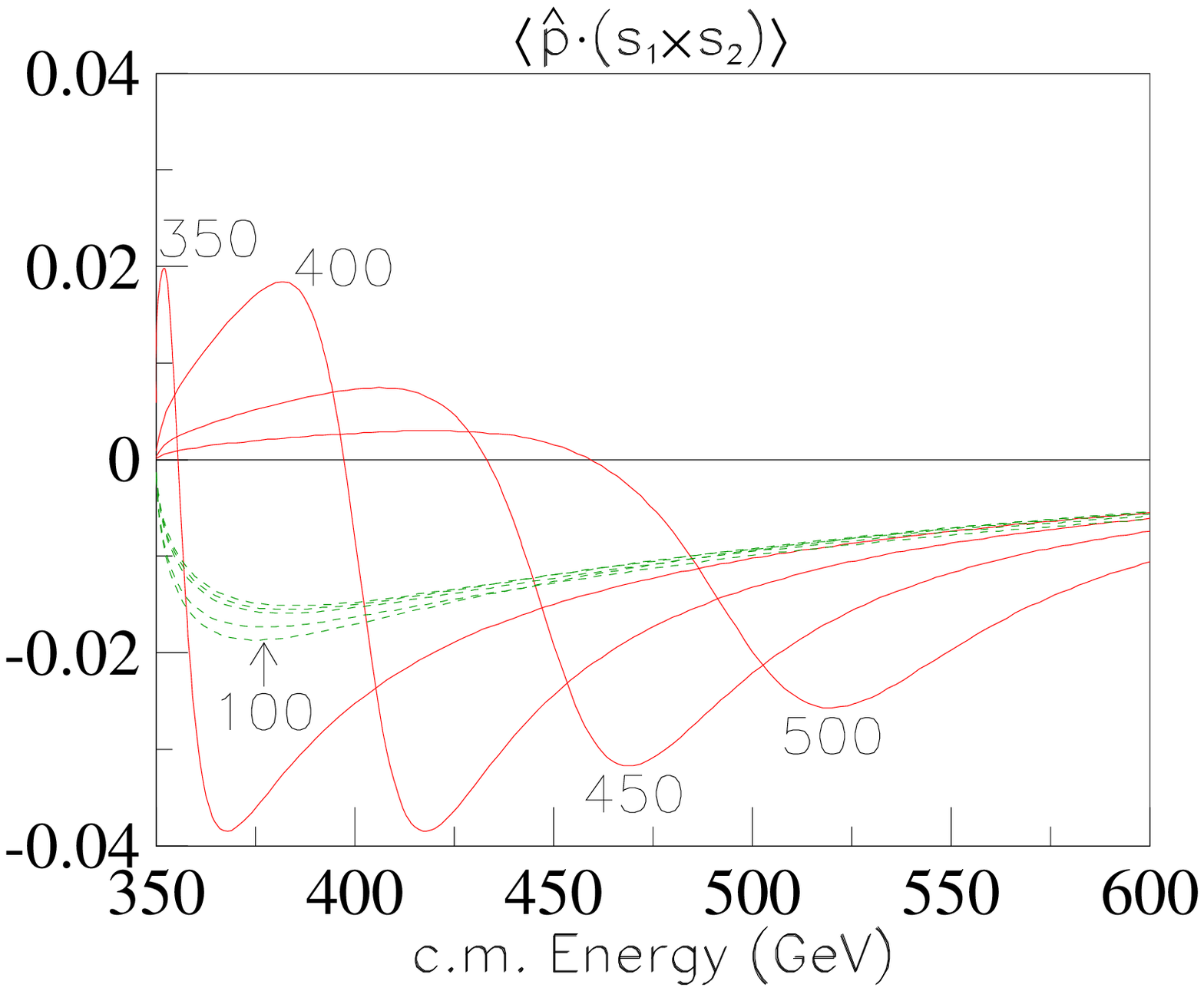}}}
\end{picture}
%\vspace*{-4mm}
\caption{Parton-level spin-spin correlations (\ref{Eq:z*bg1+bg2}) and
(\ref{Eq:z*cg1+cg2}) in $gg\to t\bar t$ {\it vs.}\ $\sqrt{\hat s}$, for
$\gamma_{CP}=1$ and different Higgs masses. Dashed: $m_H=100$, 150, \ldots,
300~GeV; solid: 350, \ldots, 500~GeV.}
\end{center}
\end{figure}
%%%%%%%%%%%%%%%%%%%%%%%%%%%%%%%%%%%%%%%%%%%%%%%%%%%%%%%%%%%%%%%%%%%%%%

We confirm the results of \cite{Bernreuther:1994hq} for (\ref{Eq:z*bg1+bg2})
(fig.~\ref{Fig:bb-cc}, left panel), but differ from their results
for (\ref{Eq:z*cg1+cg2}) (fig.~\ref{Fig:bb-cc}, right panel), since
there were sign misprints in the structure functions
$c^{(f)}_{g_1}$ and $c^{(f)}_{g_2}$ and a wrong factor of
$8\pi$ in the Higgs width $\Gamma_Z$ used in \cite{Bernreuther:1994hq}
and corrected in \cite{Bernreuther:1998qv}.
In our numerical work, we used the ``LoopTools'' package
\cite{Hahn:1998yk,vanOldenborgh:1989wn}.

As stressed in \cite{Bernreuther:1998qv}, for Higgs masses above the $t\bar t$
threshold, the quantity (\ref{Eq:z*bg1+bg2}) (and to a lesser extent
(\ref{Eq:z*cg1+cg2})) has a characteristic ``resonance--interference'' shape,
with peaks in the region $\sqrt{\hat s}\sim m_H$.  As one goes from
$\sqrt{\hat s}\lsim m_H$ to $\sqrt{\hat s}\gsim m_H$ the expectation values
(\ref{Eq:z*bg1+bg2}) and (\ref{Eq:z*cg1+cg2}) change sign. Thus, there are
cancellations which tend to reduce the CP-violating effects, when folded with
the gluon distribution functions, i.e., when integrated over the $t\bar t$
invariant mass, $M_{t\bar t}=\sqrt{\hat s}$.
%%%%%%%%%%%%%%%%%%%%%%%%%%%%%%%%%%%%%%%%%%%%%%%%%%%%%%%%%%%%%%%%%%%%%%%%
\subsection{Observables in \boldmath $pp\to t\bar t X\to l^+l^- X$}
\label{Sec:observables-pp}
%%%%%%%%%%%%%%%%%%%%%%%%%%%%%%%%%%%%%%%%%%%%%%%%%%%%%%%%%%%%%%%%%%%%%%%%

We consider events where the $t$ and/or the $\bar t$ quarks decay
semileptonically:
\begin{equation}
t \to l^+ \nu_l b, \quad \bar t \to l^- \bar\nu_l \bar b,
\end{equation}
and denote the lepton laboratory-frame momenta and energies by ${\bf l}^{\pm}$
and $E_\pm$. Following ref.~\cite{Bernreuther:1994hq}, we neglect any $CP$
violation in the top decay. This effect has been found to be small
\cite{Grzadkowski:1993gh,Grzadkowski:2000xs}.

For reference, we shall consider the observables \cite{Bernreuther:1994hq}
\begin{align} \label{Eq:A_1}
A_1&=E_+-E_-, \\
A_2&={\bf p}_{\bar t}\cdot {\bf l}^+ -{\bf p}_{t}\cdot {\bf l}^-,
\label{Eq:A_2}
\end{align}
where ${\bf p}_{t}$ (${\bf p}_{\bar t}$) is the top (anti-top) three-momentum
in the laboratory frame. The lepton energy asymmetry, $A_1$, requires events
where {\it both} the top and the anti-top decay semileptonically. The
observable $A_2$ requires the top or antitop to decay hadronically, while the
other decays semileptonically.  From the two-sample scenario discussed in
\cite{Bernreuther:1998qv} with sample ${\cal A}$ consisting of events where
the $t$ decays semileptonically and the $\bar t$ decays hadronically and
sample $\bar{\cal A}$ where the $t$ decays hadronically and the $\bar t$
decays semileptonically, one defines the expectation value of the asymmetry
$A_2$ as
\begin{align} \label{Eq:A_2_exp}
\langle A_2\rangle &=
\langle {\bf p}_{\bar t}\cdot {\bf l}^+\rangle_{{\cal A}}
-\langle {\bf p}_{t}\cdot {\bf l}^-\rangle_{\bar{\cal A}}.
\end{align}

In fig.~\ref{Fig:a1} we show the
expectation value $\langle A_1\rangle$, together with the ratio\footnote{We
shall loosely refer to this as ``signal-to-noise ratio''
\cite{Bernreuther:1994hq}.}
\begin{equation}  \label{Eq:S/N}
\frac{S}{N}=\frac{\langle A_1\rangle}
{\sqrt{\langle A_1^2\rangle-\langle A_1\rangle^2}}
\end{equation}
both as functions of the Higgs mass, and for $\gamma_{CP}=1$.  The right panel
here corresponds to fig.~7 of ref.~\cite{Bernreuther:1994hq}, but for
$m_t=175~\text{GeV}$.\footnote{We reproduce their results for the same choice
of $m_t$ and the same Higgs widths.  Also, we note that a factor of $\beta$ is
missing in eq.~(4.4) of \cite{Bernreuther:1994hq}.}  Here, and throughout this
paper, we have used the {\tt CTEQ6} parton distribution functions
\cite{Pumplin:2002vw}.  The fact that $S/N={\cal O}(10^{-3})$ means that of
the order of a million $t\bar t$ events are required (in the dilepton
channel).  Also, it means that the relative lepton energy must be known at (or
better than) the per mille level.

%%%%%%%%%%%%%%%%%%%%%%%%%%%%%%%%%%%%%%%%%%%%%%%%%%%%%%%%%%%%%%%%%%%%%%%%
\begin{figure}[htb]
\refstepcounter{figure}
\label{Fig:a1}
\addtocounter{figure}{-1}
\begin{center}
\setlength{\unitlength}{1cm}
\begin{picture}(10.0,7.0)
\put(-3.5,0)
{\mbox{\epsfysize=7.5cm
\epsffile{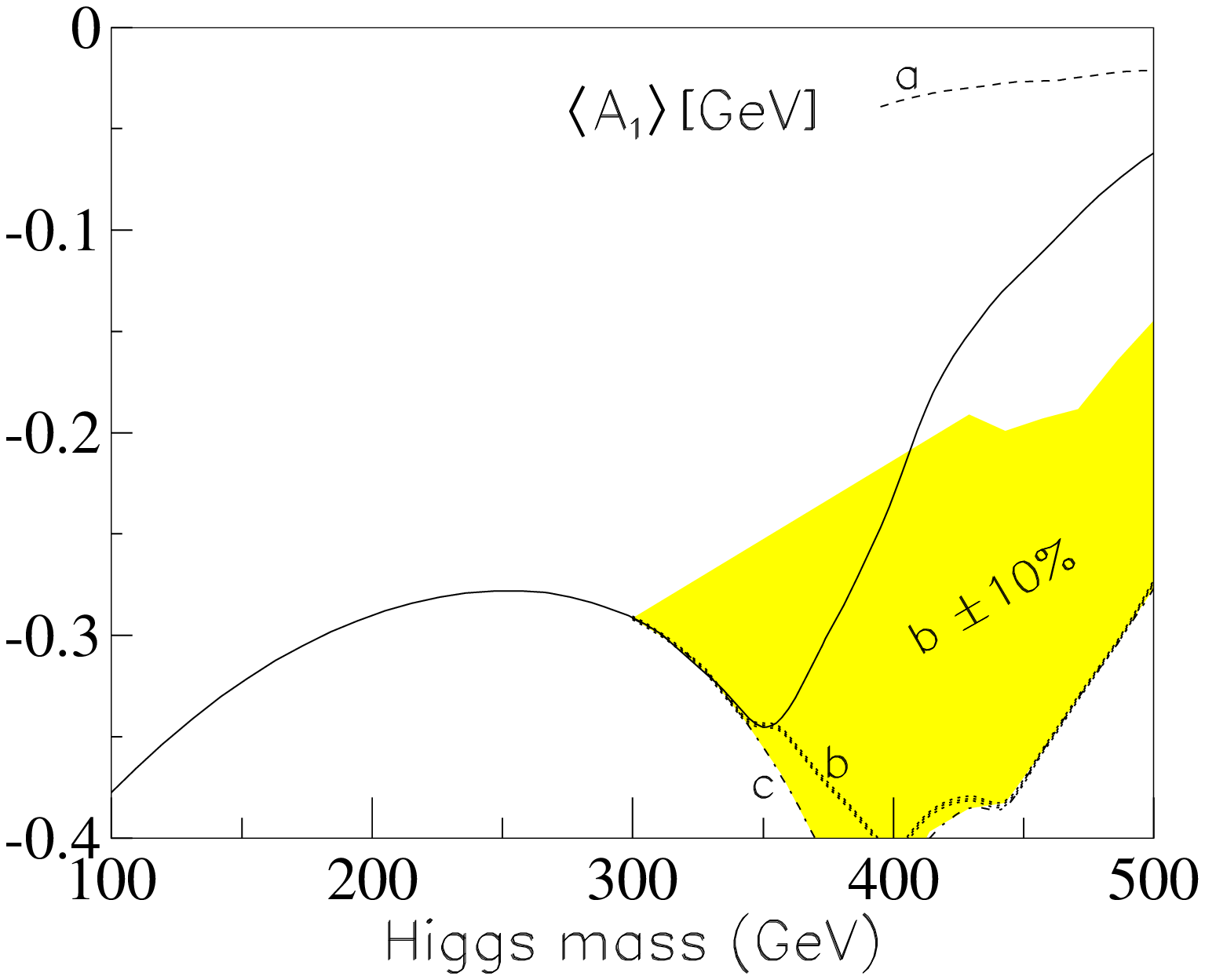}}
 \mbox{\epsfysize=7.5cm
\epsffile{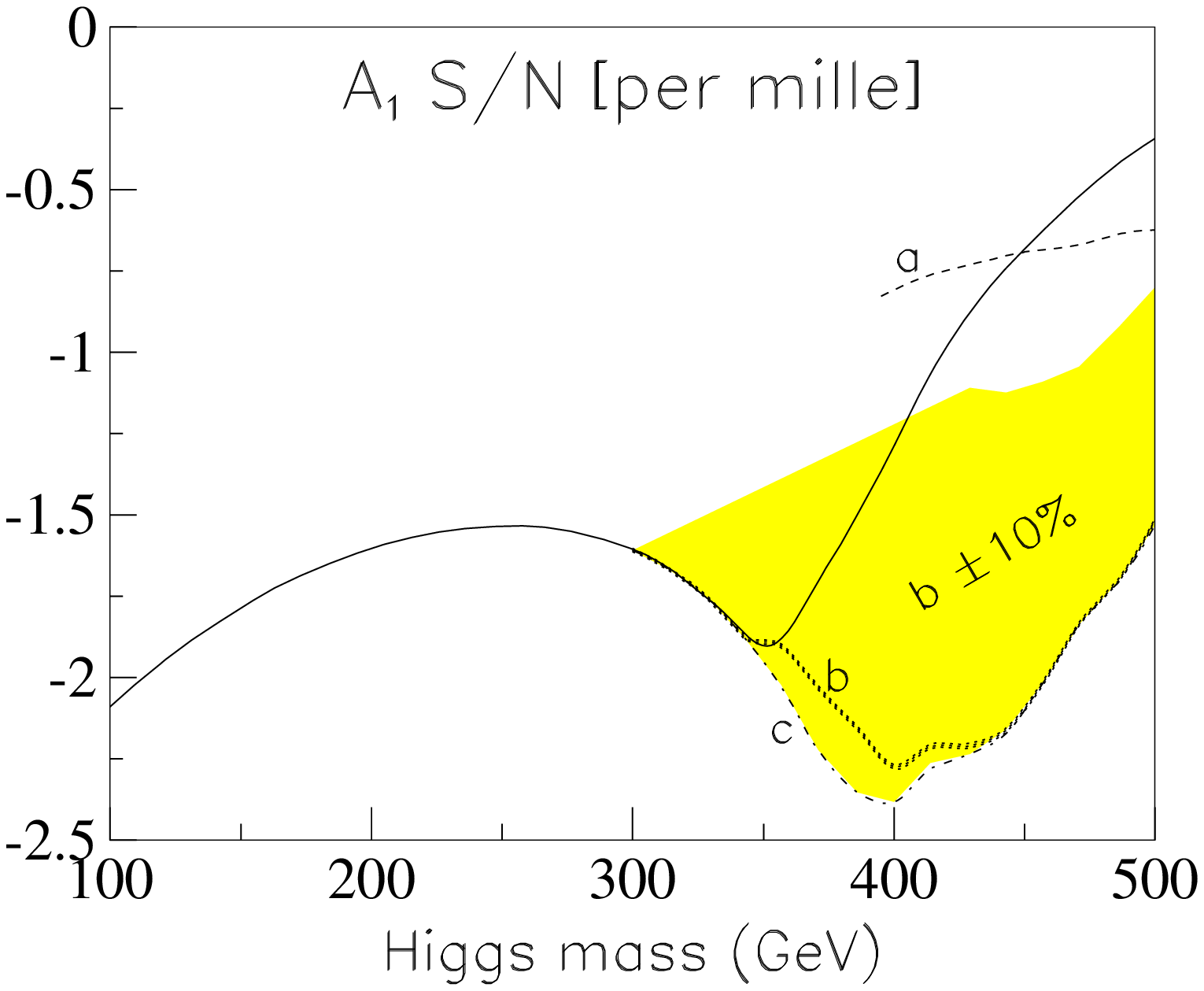}}}
\end{picture}
%\vspace*{-4mm}
\caption{{\it Left panel:} Lepton energy correlations $\langle A_1\rangle$ in
$pp\to t\bar tX$ {\it vs.}\ Higgs mass for $\gamma_{CP}=1$.  Dashed curves
labeled ``a'', ``b'' and ``c'' refer to modified observables of
eq.~(\ref{Eq:A1-modified}).  Grey (yellow) band: $\langle A_1^{(b)}\rangle$,
with 10\% uncertainty in $M_{t\bar t}$.  {\it Right panel:} Corresponding
signal-to-noise ratio.}
\end{center}
\end{figure}
%%%%%%%%%%%%%%%%%%%%%%%%%%%%%%%%%%%%%%%%%%%%%%%%%%%%%%%%%%%%%%%%%%%%%%

Next, we turn to the observable $A_2$.
We reproduce the analytical result of \cite{Bernreuther:1994hq},
their eq.~(4.5), but our numerical result, shown in fig.~\ref{Fig:a2},
has the opposite sign, and the signal-to-noise ratio is in magnitude
smaller than theirs by a factor of ${\cal O}(1/10)$ \cite{Wafaa}.
Thus, the observable $A_2$ may be rather hard to access at the LHC.

%%%%%%%%%%%%%%%%%%%%%%%%%%%%%%%%%%%%%%%%%%%%%%%%%%%%%%%%%%%%%%%%%%%%%%%%
\begin{figure}[htb]
\refstepcounter{figure}
\label{Fig:a2}
\addtocounter{figure}{-1}
\begin{center}
\setlength{\unitlength}{1cm}
\begin{picture}(10.0,7.0)
\put(-3.5,0)
{\mbox{\epsfysize=7.5cm
\epsffile{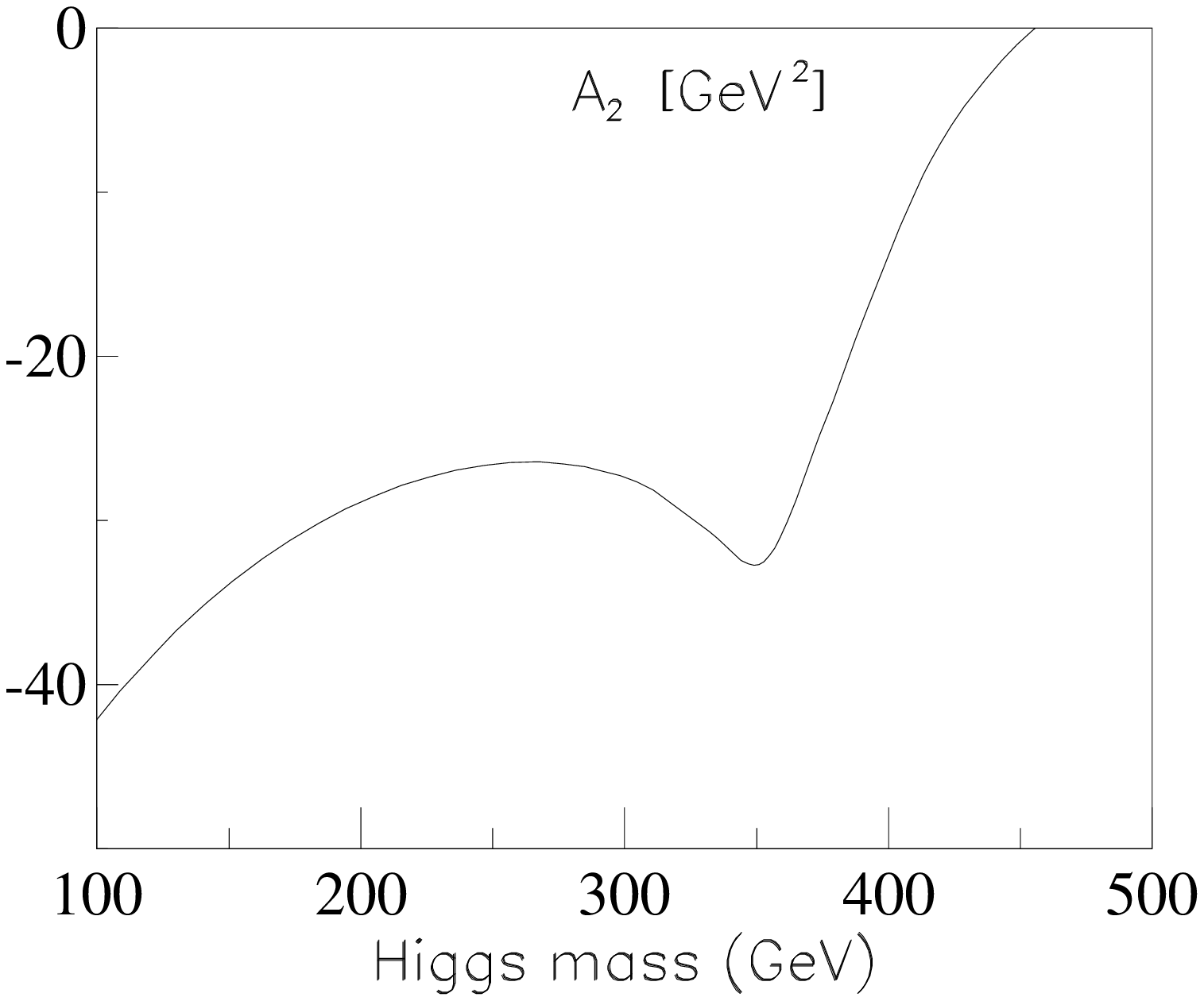}}
 \mbox{\epsfysize=7.5cm
\epsffile{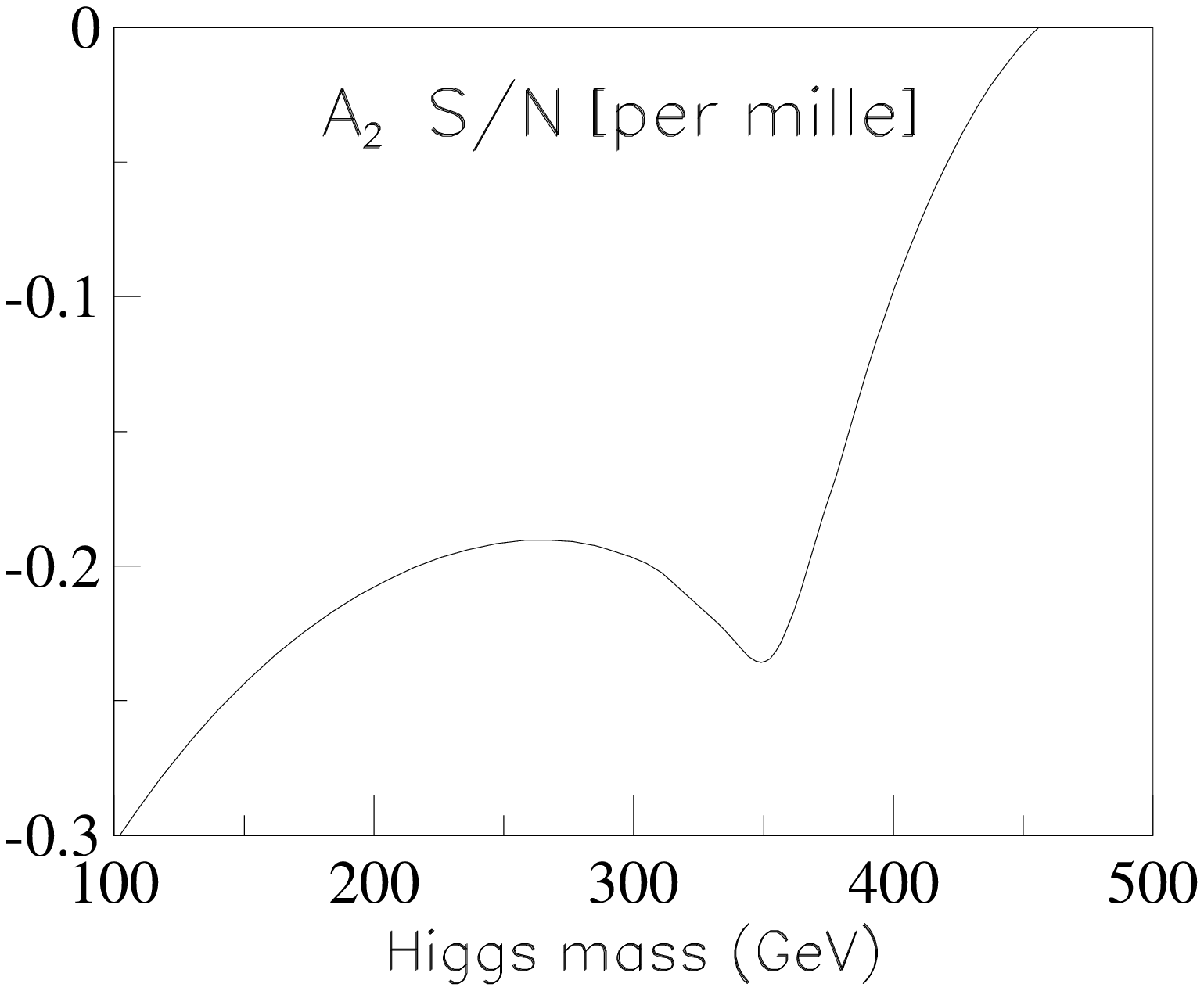}}}
\end{picture}
%\vspace*{-4mm}
\caption{{\it Left panel:} Momentum--momentum correlations $\langle
A_2\rangle$ (see eq.~(\ref{Eq:A_2})) in $pp\to t\bar tX$ {\it vs.}\ Higgs mass
for $\gamma_{CP}=1$.  {\it Right panel:} Corresponding signal-to-noise ratio.}
\end{center}
\end{figure}
%%%%%%%%%%%%%%%%%%%%%%%%%%%%%%%%%%%%%%%%%%%%%%%%%%%%%%%%%%%%%%%%%%%%%%

The reason for the small signal-to-noise ratio for $A_2$ (as compared to
$A_1$) appears to be a combination of two effects.  The main effect is due to
the Lorentz transformations.  Since it is bilinear in momenta, factors like
$B_2=1/(1-\beta^2_{gg})$ and $B_2'=\beta^2_{gg}/(1-\beta^2_{gg})$ are
involved, {\it vs.}\ $B_1=1/\sqrt{1-\beta^2_{gg}}$ for $A_1$, with
$\beta_{gg}$ the gluon--gluon c.m.\ velocity w.r.t.\ the laboratory frame. The
``signal-to-noise'' ratios for these quantities (``quasi-observables'') $B_2$
and $B_2'$ are reduced with respect to that of $B_1$ by a factor of 3--4. This
is due to the importance of events where the $t\bar t$ invariant mass is close
to threshold, and the relatively light $t\bar t$ system has a high velocity in
the laboratory system.  Furthermore, in $A_2$, higher powers of cosines of
angles are involved, and we note that, for example,
$\langle\cos^2\theta\rangle=1/3$, whereas $\langle\cos^4\theta\rangle=1/5$.
Together, these two effects give a reduction by ${\cal O}(1/10)$.

For values of the Higgs mass above the $t\bar t$ threshold, there is a
fall-off in the absolute value of $S/N$. This is in part due to cancellation
between the positive and negative parts in $\langle\hat{\bf p}\cdot({\bf
s}_1-{\bf s}_2)\rangle$, as shown in fig.~\ref{Fig:bb-cc}.  One possible way
to enhance the signal, would be to consider bins in $M_{t\bar t}$, centered
around the resonant contribution of the (lightest) Higgs particle
\cite{Bernreuther:1998qv}.  
It would be worthwhile to perform a detailed simulation study of the gains
of such binning in $M_{t\bar t}$. The value of this technique would depend
critically on the mass resolution that can be achieved by a given detector.

We have investigated various other ways to reduce this loss of sensitivity, by
studying instead of $A_1$, the modified observables (still for events with
two leptons):
\begin{align} \label{Eq:A1-modified}
A_1^{(a)}&=\frac{M_{t\bar t}-M_H}{M_{t\bar t}+M_H}\,A_1, \\
A_1^{(b)}&={\rm sign}(M_{t\bar t}-M_H)\,A_1, \\
A_1^{(c)}&=-|A_1|.
\end{align}
The expectation values and signal-to-noise ratios for these observables are
also shown in fig.~\ref{Fig:a1}, labeled ``a'', ``b'' and ``c''. For
$A_1^{(b)}$, and for high Higgs masses, a significant improvement is obtained.
However, the modified observables $A_1^{(a)}$ and $A_1^{(b)}$ require precise
knowledge of the $t\bar t$ invariant mass, as well as of the Higgs mass.  Of
these, presumably the $t\bar t$ invariant mass will be the most difficult one
to determine. We show in fig.~\ref{Fig:a1} the ranges in $\langle
A_1^{(b)}\rangle$ and signal-to-noise ratio that result from errors of 10\% in
this invariant mass.

If only the magnitude of $A_1$ is required, then it can be determined with
better sensitivity from $A_1^{(c)}=-|A_1|$, which in fig.~\ref{Fig:a1}
coincides with the lower boundary of the shaded region and hence gives
the largest signal enhancement.

%%%%%%%%%%%%%%%%%%%%%%%%%%%%%%%%%%%%%%%%%%%%%%%%%%%%%%%%%%%%%%%%%%%%%%%%
\section{The Two-Higgs-Doublet Model}
\setcounter{equation}{0}
%%%%%%%%%%%%%%%%%%%%%%%%%%%%%%%%%%%%%%%%%%%%%%%%%%%%%%%%%%%%%%%%%%%%%%%%
In the previous discussion, the amount of $CP$ violation was given by the
Yukawa couplings $a$, $\tilde a$ (in particular, via the product,
$\gamma_{CP}=-a\tilde a$) and the Higgs mass.  In an explicit model, the
situation is more complex, since there are several Higgs bosons, whose masses
and couplings will be inter-related by the specific model.  As an example of
such a more complex situation, we shall in this and the next section study a
specific Two-Higgs-Doublet Model where the $CP$ violation is {\it minimal}
in structure.
%%%%%%%%%%%%%%%%%%%%%%%%%%%%%%%%%%%%%%%%%%%%%%%%%%%%%%%%%%%%%%%%%%%%%%%%
\subsection{Parametrization} \label{sect-parametrization}
%%%%%%%%%%%%%%%%%%%%%%%%%%%%%%%%%%%%%%%%%%%%%%%%%%%%%%%%%%%%%%%%%%%%%%%%

The Two-Higgs-Doublet Model we want to consider, is the one
discussed in \cite{Ginzburg:2001ss},
where the potential is given by
\begin{eqnarray}                    \label{Eq:gko-pot}
V&=&\frac{\lambda_1}{2}(\phi_1^\dagger\phi_1)^2
+\frac{\lambda_2}{2}(\phi_2^\dagger\phi_2)^2
+\lambda_3(\phi_1^\dagger\phi_1) (\phi_2^\dagger\phi_2)
+\lambda_4(\phi_1^\dagger\phi_2) (\phi_2^\dagger\phi_1) \\
&&+\frac{1}{2}\left[\lambda_5(\phi_1^\dagger\phi_2)^2+{\rm h.c.}\right]
-\frac{1}{2}\left\{m_{11}^2(\phi_1^\dagger\phi_1)
+\left[m_{12}^2 (\phi_1^\dagger\phi_2)+{\rm h.c.}\right]
+m_{22}^2(\phi_2^\dagger\phi_2)\right\}. \nonumber 
\end{eqnarray}
The parameters $\lambda_5$ and $m_{12}^2$ are allowed to be complex,
subject to the constraint
\begin{equation}
\Im m_{12}^2=\Im \lambda_5\, v_1 v_2,
\end{equation}
with $v_1$ and $v_2$ the vacuum expectation values ($v_1^2+v_2^2=v^2$,
with $v=246\text{ GeV}$).
It is this quantity, $\Im m_{12}^2$ (or $\Im \lambda_5$) which leads
to $CP$ violation, and one may think of $CP$ violation as being introduced
softly, via the mass term $m_{12}^2$ in (\ref{Eq:gko-pot}).
Since the potential has a $Z_2$ symmetry that is only broken softly
by the $m_{12}^2$ term, flavour-changing neutral currents
are suppressed \cite{Glashow:1976nt,Branco}.

The neutral-sector mass squared matrix corresponding to the potential
(\ref{Eq:gko-pot}) can be written as (for details, see \cite{Ginzburg:2001ss})
\begin{equation}          \label{Eq:MM}
{\cal M}=v^2
\begin{pmatrix}
\lambda_1c_\beta^2+\nu s_\beta^2&
(\lambda_{345} -\nu)c_\beta s_\beta&
-\half\Im\lambda_5\, s_\beta\\[4mm] 
(\lambda_{345} -\nu)c_\beta s_\beta&
\lambda_2s_\beta^2+\nu c_\beta^2&
-\half\Im\lambda_5\, c_\beta\\[4mm] 
-\half\Im\lambda_5\, s_\beta& 
-\half\Im\lambda_5\, c_\beta&
-\Re\,\lambda_5+\nu
\end{pmatrix}
\end{equation}
with the abbreviations $c_\beta=\cos\beta$, $s_\beta=\sin\beta$,
$\tan\beta=v_2/v_1$, $\lambda_{345}=\lambda_3+\lambda_4+\Re\lambda_5$,
$\nu=\Re m_{12}^2/(2v^2\sin\beta\cos\beta)$ and $\mu^2=v^2\nu$.

For $\Im\lambda_5\ne0$, the elements ${\cal M}_{13}$ and ${\cal M}_{23}$
provide mixing between the $2\times2$ upper left part of the matrix and ${\cal
M}_{33}$.  These two sectors would otherwise represent two scalar
(usually denoted $h$ and $H$) and one pseudoscalar Higgs boson (denoted $A$).
We note that these two ``mixing elements'' are related via $\tan\beta$:
\begin{equation}  \label{Eq:M13-M23}
{\cal M}_{13}=\tan\beta\, {\cal M}_{23}.
\end{equation}
It is in this sense that the $CP$ violation is {\it minimal} in structure.
This simple relation is violated in more general models, with
so-called $\lambda_6$ and $\lambda_7$ terms in the potential
\cite{Ginzburg:2001ss}.

We diagonalize (\ref{Eq:MM}) with the matrix $R$, defined such that
\begin{equation}
R{\cal M}R^{\rm T}={\rm diag}(M_1^2,M_2^2,M_3^2),
\end{equation}
and use for the rotation matrix \cite{Groom:in}
\begin{align}     \label{Eq:R-angles}
R=R_c\,R_b\,R_{\tilde\alpha}
=&\begin{pmatrix}
1         &    0         &    0 \\
0 &  \cos\alpha_c & \sin\alpha_c \\
0 & -\sin\alpha_c & \cos\alpha_c
\end{pmatrix}
\begin{pmatrix}
\cos\alpha_b & 0 & \sin\alpha_b \\
0         &       1         & 0 \\
-\sin\alpha_b & 0 & \cos\alpha_b
\end{pmatrix}
\begin{pmatrix}
\cos\tilde\alpha & \sin\tilde\alpha & 0 \\
-\sin\tilde\alpha & \cos\tilde\alpha & 0 \\
0         &       0         & 1
\end{pmatrix}  \nonumber \\
=&\begin{pmatrix}
c_{\tilde\alpha}\,c_b & s_{\tilde\alpha}\,c_b & s_b \\
- (c_{\tilde\alpha}\,s_b\,s_c + s_{\tilde\alpha}\,c_c) 
& c_{\tilde\alpha}\,c_c - s_{\tilde\alpha}\,s_b\,s_c & c_b\,s_c \\
- c_{\tilde\alpha}\,s_b\,c_c + s_{\tilde\alpha}\,s_c 
& - (c_{\tilde\alpha}\,s_c + s_{\tilde\alpha}\,s_b\,c_c) & c_b\,c_c
\end{pmatrix}
\end{align}
with $c_i=\cos\alpha_i$, $s_i=\sin\alpha_i$.
The angular range, beyond which $R$ is repeated, can be chosen as
$-\pi/2<\tilde\alpha\le\pi/2$, $-\pi<\alpha_b\le\pi$, and
$-\pi/2<\alpha_c\le\pi/2$.
However, the physical range is more restricted, since we require
$M_1\le M_2\le M_3$.

There are certain symmetries, under which the masses are unchanged, but one or
two rows of $R$ (i.e., physical Higgs fields) change sign. 
These symmetries are:
\begin{alignat}{5}  \label{Eq:symm-A-B}
&\text{A}:&\quad 
&\tilde\alpha\text{ fixed},\;\alpha_c\to-\alpha_c,\;
\alpha_b\to\pi+\alpha_b: &\quad
R_{1i}&\to -R_{1i}, &\quad R_{2i}&\to R_{2i}, &\quad R_{3i}&\to -R_{3i},
\nonumber \\
&\text{B}:&\quad 
&\tilde\alpha\to\pi+\tilde\alpha,\; \alpha_b\to-\alpha_b,\;
\alpha_c\to-\alpha_c: &\quad
R_{1i}&\to -R_{1i}, &\quad R_{2i}&\to -R_{2i}, &\quad R_{3i}&\to R_{3i}.
\end{alignat}
In addition, we have the following symmetry for 
$\tilde\alpha$, $\alpha_c$ fixed:
\begin{equation}
\text{C}:\quad 
\alpha_b\to\pi-\alpha_b,\; \alpha_b>0\;
(\alpha_b\to-\pi-\alpha_b,\; \alpha_b<0):\quad
R_{1i}\to -R_{1i}, \quad R_{j3}\to -R_{j3},
\end{equation}
the other elements in $R$ being unchanged.  
Finally, we have, for $\alpha_b$, $\alpha_c$ fixed:
\begin{equation}
\text{D}:\quad 
\tilde\alpha\to\pi+\tilde\alpha:\quad
R_{j1}\to -R_{j1}, \quad R_{j2}\to -R_{j2}, \quad R_{j3}\to R_{j3}.
\end{equation}
(The symmetries B and D are of marginal interest, since they only relate the
{\it edges} of the angular range.)

The $CP$-conserving case is obtained by taking $\alpha_b=0$ or $\alpha_b=\pi$,
together with $\alpha_c=0$ and $\tilde\alpha=\half\pi+\alpha$ arbitrary. Here,
$\alpha$ is the familiar mixing angle of the $CP$-even sector. Thus, as
alternatives to $\Im\lambda_5$, the angles $\alpha_b$ and $\alpha_c$
parametrize the mixing that leads to $CP$ violation. Of course, replacing one
parameter by two, constraints are imposed on the mass spectrum.

%%%%%%%%%%%%%%%%%%%%%%%%%%%%%%%%%%%%%%%%%%%%%%%%%%%%%%%%%%%%%%%%%%%%%%%%%%%%%%
\subsubsection*{Yukawa couplings}
%%%%%%%%%%%%%%%%%%%%%%%%%%%%%%%%%%%%%%%%%%%%%%%%%%%%%%%%%%%%%%%%%%%%%%%%%%%%%%

With this notation, and adopting the so-called Model~II \cite{HHG} for the
Yukawa couplings, where the down-type and up-type quarks are coupled only
to $\phi_1$ and $\phi_2$, respectively, the $Ht\bar t$ couplings can be
expressed (relative to the SM coupling) as
\begin{equation}  \label{Eq:H_itt}
H_j t \bar t: \qquad 
\frac{1}{\sin\beta}\, [R_{j2}-i\gamma_5\cos\beta R_{j3}]
\equiv a+i\tilde a\gamma_5.
\end{equation}

As mentioned in the Introduction, the product of the scalar and pseudoscalar
couplings,
\begin{equation}   \label{Eq:gamma_CP-i}
\gamma_{CP}=-a\,\tilde a
=\frac{\cos\beta}{\sin^2\beta}\, R_{j2}R_{j3}
\end{equation}
plays an important role in determining the amount of $CP$ violation.
We note the following symmetries of $\gamma_{CP}$ for any of the three
Higgs bosons $H_j$ ($j=1,2,3$):
\begin{itemize}
\item
under the symmetries ``A'' and ``B'' of (\ref{Eq:symm-A-B}),
$\gamma_{CP}$ is invariant,
\item
under the symmetries ``C'' and ``D'', $\gamma_{CP}$ changes sign.
\end{itemize}

As was seen in sect.~\ref{Sec:model-indep},
unless the Higgs boson is resonant with the $t\bar t$ system, $CP$ violation
is largest for small Higgs masses. We shall therefore focus on the
contributions of the lightest Higgs boson, $H_1$.
For the lightest Higgs boson, the coupling (\ref{Eq:H_itt}) becomes
\begin{equation}    \label{Eq:gamma_CP}
H_1 t \bar t: \qquad 
\frac{1}{\sin\beta}\,[\sin\tilde\alpha\cos(\alpha_b) 
                   -i\gamma_5\cos\beta\sin(\alpha_b)], \quad
\text{with} \quad 
\gamma_{CP}=\half\,\frac{\sin\tilde\alpha\sin(2\alpha_b)}{\tan\beta\sin\beta},
\end{equation}
where $\tilde\alpha$ and $\alpha_b$ are 
mixing angles of the Higgs mass matrix as defined above.

When the three neutral Higgs bosons are light, they will {\it all} contribute
to the $CP$-violating effects. In fact, in the limit of three mass-degenerate
Higgs bosons, the $CP$ violation will cancel, since [cf.\
eq.~(\ref{Eq:gamma_CP-i})]
\begin{equation}  \label{Eq:orthogonal}
\sum_{j=1}^3\gamma_{CP}
=\frac{\cos\beta}{\sin^2\beta}\sum_{j=1}^3 R_{j2}\,R_{j3}=0
\end{equation}
due to the orthogonality of $R$. 

%%%%%%%%%%%%%%%%%%%%%%%%%%%%%%%%%%%%%%%%%%%%%%%%%%%%%%%%%%%%%%%%%%%%%%%%%%%%%%
\subsubsection*{\boldmath ``Maximal $CP$ violation''}
%%%%%%%%%%%%%%%%%%%%%%%%%%%%%%%%%%%%%%%%%%%%%%%%%%%%%%%%%%%%%%%%%%%%%%%%%%%%%%

The $|\gamma_{CP}|$ of eq.~(\ref{Eq:gamma_CP}) is readily maximized w.r.t.\
the rotation angles $\tilde\alpha$ and $\alpha_b$:
\begin{equation}   \label{Eq:max_CP}
\tilde\alpha^{{\rm max}\, CP}=\pm\half\pi,\quad
\alpha_b^{{\rm max}\, CP}=\pm\fourth\pi\ \text{ or }\ \pm\threefourth\pi,
\end{equation}
whereas it increases with {\it decreasing} values of $\tan\beta$.  Negative
values of $\alpha_c$ are related to positive values through the symmetry ``A''
of eq.~(\ref{Eq:symm-A-B}) and need not be considered separately.  
Similarly, the case of $\alpha_b=3\pi/4$ is related to that of 
$\alpha_b=\pi/4$ by the symmetry ``C'' and just leads to an over-all
change of sign for $\gamma_{CP}$. We shall
somewhat imprecisely (since also $H_2$ and $H_3$ contribute) refer to these
cases (\ref{Eq:max_CP}) as ``maximal $CP$ violation''.

More generally, large $CP$-violating effects are expected when 
$|\Im\lambda_5|$ is large, or, when the
elements ${\cal M}_{13}$ and ${\cal M}_{23}$ are large, see
(\ref{Eq:MM}). These can be expressed in terms of the rotation matrix
and the mass eigenvalues as
\begin{align} \label{Eq:max-mix}
{\cal M}_{13}&=R_{11}R_{13}M_1^2 + R_{21}R_{23}M_2^2 
+ R_{31}R_{33}M_3^2, \nonumber \\
{\cal M}_{23}&=R_{12}R_{13}M_1^2 + R_{22}R_{23}M_2^2 
+ R_{32}R_{33}M_3^2.
\end{align}
In this description, one measure of ``maximal mixing'' would be
to ignore the lightest mass, and then require the coefficients
of $M_2^2$ and $M_3^2$ to be equal in magnitude.
This would require $|\sin(\tilde\alpha)|=0$ or $|\cos(\tilde\alpha)|=0$,
i.e., $\tilde\alpha=0$, $\pm\half\pi$ or $\pi$, and simultaneously
$|\sin(2\alpha_b)|=|\sin(2\alpha_c)|=1$.
We shall refer to these cases as ``maximal mixing''.

The bounds on the electric dipole moments of the neutron and the electron
impose restrictions on the allowed magnitude of $CP$ violation in the 2HDM.
We note that, within a consistent framework, and for maximal $CP$ violation in
the Higgs--gauge sector \cite{Mendez:1991gp} (which amounts to maximizing the
product of the three couplings (\ref{Eq:HZZ-couplings}), and thus is different
from the present concept of ``maximal $CP$ violation'', where we consider
Yukawa interactions only), those bounds restrict the mass splitting between
$M_2$ and $M_3$ to be less than ${\cal O}(15\%-20\%)$ \cite{Hayashi:1994ha}.

%%%%%%%%%%%%%%%%%%%%%%%%%%%%%%%%%%%%%%%%%%%%%%%%%%%%%%%%%%%%%%%%%%%%%%%%%%%%%%
\subsection{Physical content}  \label{Sec:physical-content}
%%%%%%%%%%%%%%%%%%%%%%%%%%%%%%%%%%%%%%%%%%%%%%%%%%%%%%%%%%%%%%%%%%%%%%%%%%%%%%
Specifying all the parameters of the potential, as well as the structure of
the Yukawa couplings, the physical content of the model is fixed. We shall
here follow a somewhat different approach: We start out by specifying the
masses of two Higgs bosons (the lightest and one other), together with
$\tan\beta$ and the three angles that determine $R$.  Then, unless
$\alpha_b=0$ or $\alpha_c=0$, the third Higgs mass can be determined, as well
as all the couplings.  This approach gives more ``control'' of the physical
input.

With $\tan\beta$, $R$ and the masses of {\it two} neutral Higgs bosons fixed,
the third one is given by
\begin{equation}  \label{Eq:masses}
 R_{13}(R_{11}-R_{12}\tan\beta)M_1^2
+R_{23}(R_{21}-R_{22}\tan\beta)M_2^2
+R_{33}(R_{31}-R_{32}\tan\beta)M_3^2=0,
\end{equation}
where we have assumed $\Im\lambda_5\ne0$ and made use of the relation
(\ref{Eq:M13-M23}).
Invoking the orthogonality of $R$, one sees that
this relation (\ref{Eq:masses}) only relates {\it differences}
of masses squared:
\begin{equation}   \label{Eq:masses_diff}
 R_{23}(R_{21}-R_{22}\tan\beta)(M_2^2-M_1^2)
+R_{33}(R_{31}-R_{32}\tan\beta)(M_3^2-M_1^2)=0.
\end{equation}
In general, as two Higgs masses approach each other, also the third mass has
to approach the same value, as we can read off
eq.~(\ref{Eq:masses_diff}).  However, for particular choices of the
parameters, this is not the case. For example, if we put the coefficient
of $M_1^2$ in eq.~(\ref{Eq:masses}) equal to zero, one has $M_2=M_3$ for
\begin{equation}
R_{13}(R_{11}-R_{12}\tan\beta)=0.
\end{equation}
This equation is satisfied for {\it (i)} $R_{13}=0$, or {\it (ii)}
$\tan\beta=R_{11}/R_{12}$ or {\it (iii)} $R_{11}$ and $R_{12}$ each
zero. Solution {\it (i)} implies $\alpha_b=0$, whereas solution {\it (ii)}
implies $\tan\beta=\cot\tilde\alpha$, or
$\tilde\alpha=\half\pi-\beta$. Solution{\it (iii)} implies $|\alpha_b|=\pi/2$
where, non-interestingly, the three masses are arbitrarily chosen. Following
from {\it (ii)}, for example, we get $M_2=M_3$ for arbitrary $M_1$, for the
choice of parameters $\tan\beta=1$ and $\tilde\alpha=\pi/4$.

%%%%%%%%%%%%%%%%%%%%%%%%%%%%%%%%%%%%%%%%%%%%%%%%%%%%%%%%%%%%%%%%%%%%%%%%%%%%%%
\subsubsection*{Allowed regions in the parameter space}
%%%%%%%%%%%%%%%%%%%%%%%%%%%%%%%%%%%%%%%%%%%%%%%%%%%%%%%%%%%%%%%%%%%%%%%%%%%%%%
In fig.~\ref{Fig:alphas-1} we show the allowed regions in the
$\alpha_b$--$\alpha_c$ plane for selected values of $\tan\beta$ and
$\tilde\alpha$, subject to the constraint $M_1\le M_2\le M_3$.  Regions of
negative $\alpha_c$ and $|\alpha_b|>\pi/2$ are not shown, they follow by the
symmetries ``A'' and ``C'', respectively, of eq.~(\ref{Eq:symm-A-B}).

In this analysis, we keep $M_1$ and $M_2$ fixed, and determine $M_3$ from
eq.~(\ref{Eq:masses}). Clearly, this breaks down when
\begin{equation}  \label{Eq:M3-undefined}
R_{33}(R_{31}-R_{32}\tan\beta)=0,
\end{equation}
in which case $M_3$ is not determined. 
This occurs deep inside the unphysical region, as explained below.

The dark (blue) regions in fig.~\ref{Fig:alphas-1} are allowed, the white ones
are not allowed. Some reasons why a region is physically forbidden are: $(i)$
the equation for $M_3$ has no solution (for example, when $M_3^2<0$, or when
eq.~(\ref{Eq:M3-undefined}) holds), $(ii)$ $M_3>M_2$ is violated, $(iii)$
positivity of the potential (\ref{Eq:gko-pot}) is violated, or $(iv)$
perturbativity is violated.  The requirement to perturbativity is taken as
\begin{equation}   \label{Eq:pert-xi}
|\lambda_i|<4\pi\xi,
\end{equation}
where we somewhat arbitrarily take $\xi=0.8$.  The lightly shaded (yellow)
regions are also allowed, if we use a less stringent condition on
perturbativity, $\xi=1.0$.\footnote{These numbers are of course only
order-of-magnitude indicators. For example, one could argue that since
$\lambda_1$ and $\lambda_2$ are accompanied by factors $1/2$ in the potential
(and hence in all couplings), the relevant limit is $8\pi$, not $4\pi$.}
The cut-off against the disallowed region depends
on the additional parameters, which we in this figure take as
$M_1=100~\text{GeV}$, $M_2=300~\text{GeV}$, $M_{H^\pm}=500~\text{GeV}$, and
$\mu=0$ (corresponding to Case~I, studied in
sect.~\ref{Sec:higgs-light-light-heavy}).  This dependence enters via the
conversion of the masses of the neutral Higgs bosons, and the rotation angles,
to $\lambda$'s.
%%%%%%%%%%%%%%%%%%%%%%%%%%%%%%%%%%%%%%%%%%%%%%%%%%%%%%%%%%%%%%%%%%%%%%%%
\begin{figure}[htb]
\refstepcounter{figure}
\label{Fig:alphas-1}
\addtocounter{figure}{-1}
\begin{center}
\setlength{\unitlength}{1cm}
\begin{picture}(15.0,12.0)
\put(0,-0.7)
{\mbox{\epsfysize=13cm
 \epsffile{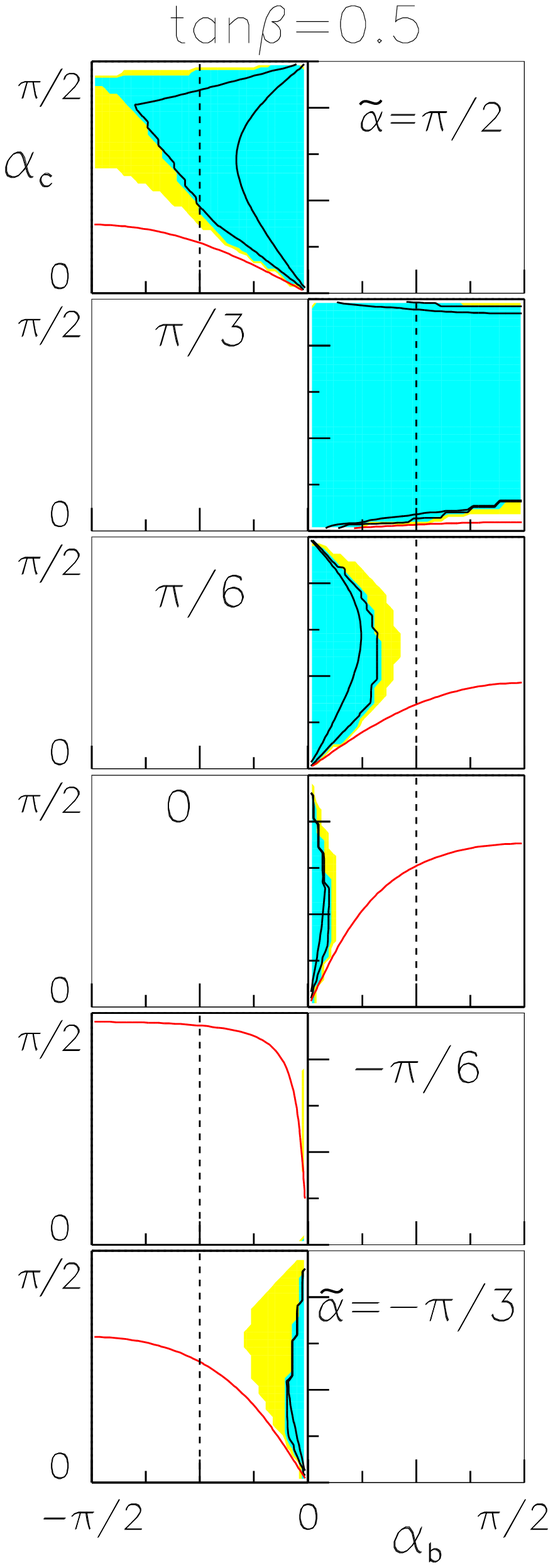}}
 \mbox{\epsfysize=13cm
 \epsffile{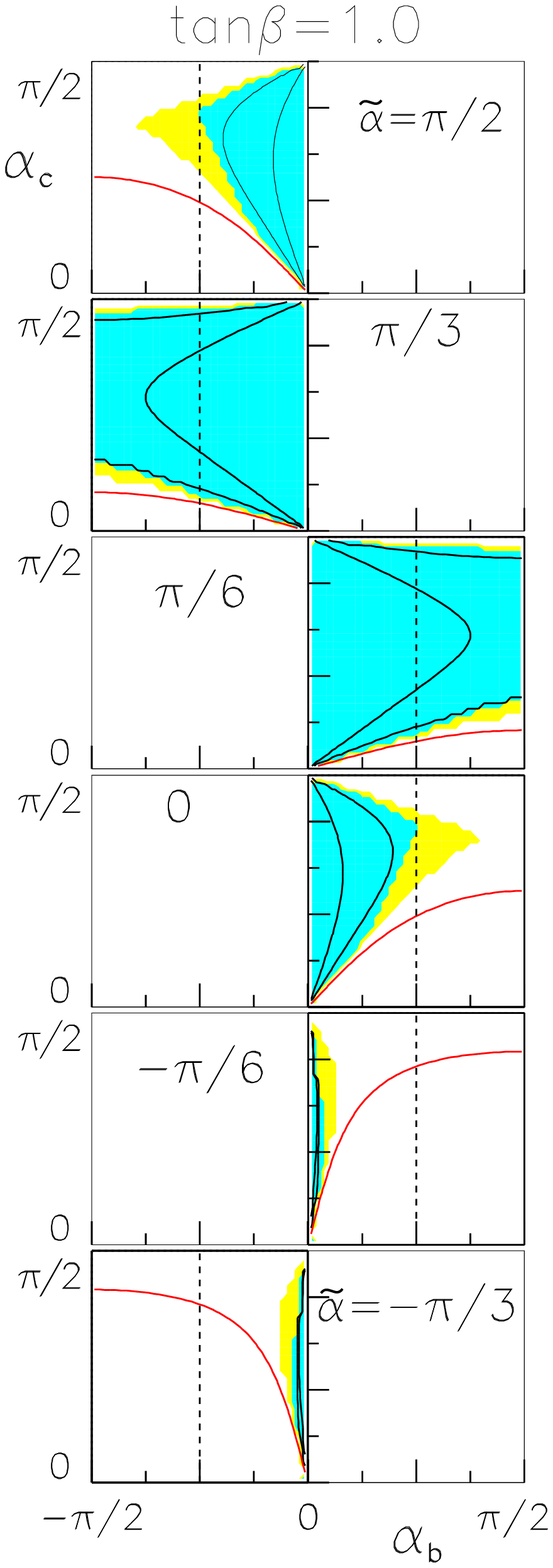}}
 \mbox{\epsfysize=13cm
\epsffile{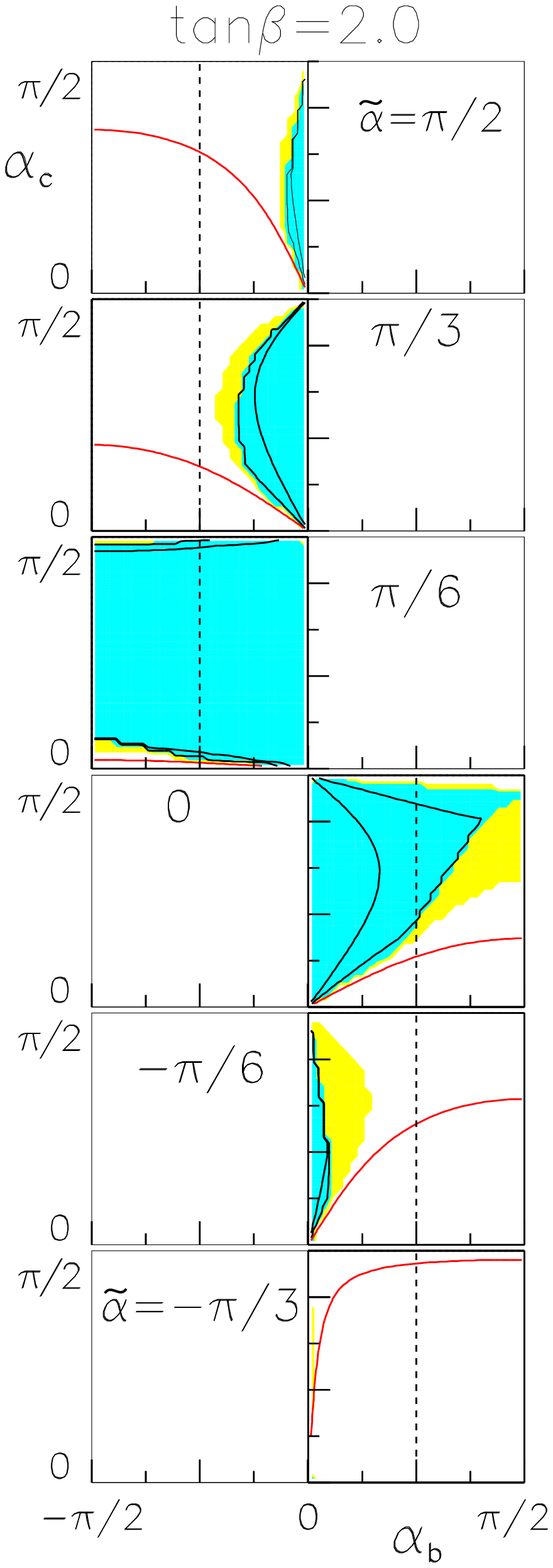}}}
\end{picture}
%\vspace*{-4mm}
\caption{Physically allowed regions in the $\alpha_b$--$\alpha_c$ plane, for
$\tan\beta=0.5$, 1 and 2 and for selected values of $\tilde\alpha$.  Blue
(dark): allowed regions; white: not allowed; yellow: marginally allowed (see
text).  The dashed lines at $\alpha_b=\pm\pi/4$ are lines of ``maximum $CP$
violation''.  Solid contours in allowed regions: $M_2/M_3=0.6$ and 0.8
($M_2=M_3$ along $\alpha_b=0$).  For no choice of mass parameters can the
allowed region extend beyond the solid contours in the forbidden region.}
\end{center}
\end{figure}
%%%%%%%%%%%%%%%%%%%%%%%%%%%%%%%%%%%%%%%%%%%%%%%%%%%%%%%%%%%%%%%%%%%%%%

The condition (\ref{Eq:M3-undefined}), which is satisfied for
\begin{equation}  \label{Eq:two-cases}
(1)\quad
R_{33}=0,\quad\text{ or}\quad
(2)\quad
R_{31}-R_{32}\tan\beta=0,
\end{equation}
defines borders where $M_3^2$ changes sign.  We note that {\it these borders
are independent of the mass parameters of the model}. Thus, they are absolute
borders which the physically allowed region can never cross for any choice of
mass parameters.  Let us now discuss the two cases of
eq.~(\ref{Eq:two-cases}).  In the $\alpha_b$--$\alpha_c$ plane, case (1) is
satisfied when $\cos\alpha_b=0$ or when $\cos\alpha_c=0$.  When
$\cos\alpha_b=0$, then eq.~(\ref{Eq:masses}) is trivially satisfied, so this
is not interesting.  The second subcase, $\cos\alpha_c=0$, corresponds to the
upper boundary of the plots in fig.~\ref{Fig:alphas-1}, $\alpha_c=\pi/2$.
Since this line represents a border where $M_3^2$ changes from $+\infty$ to
$-\infty$, there is a region next to it where $M_3$ is so large that
perturbativity breaks down (and hence forbidden), as indeed seen in the
figure.

Coming now to the more interesting case (2) of (\ref{Eq:two-cases}), 
which, according to eq.~(\ref{Eq:R-angles}) is satisfied for
\begin{equation}  \label{Eq:angular-relation}
\tan(\beta+\tilde\alpha)\,\tan\alpha_c=\sin\alpha_b,
\end{equation}
we obtain the contours shown in fig.~\ref{Fig:alphas-1} deep inside
the unphysical region.

These contours must be in the same ``quadrant'' in the $\alpha_b$--$\alpha_c$
plane as where the physically allowed region is located. This can be seen as
follows. On one side of the contour, $M_3^2$ is large and positive.  By
continuity, moving from the contour toward the line $\alpha_b=0$, where
$M_3=M_2$, the value of $M_3$ will inevitably pass through an allowed region
where $M_3\gsim{\cal O}(M_2)$.

This observation allows us to determine in which quadrant the physically
allowed regions are located. Near the origin, (\ref{Eq:angular-relation})
can be approximated as
\begin{equation}  \label{Eq:approx-angular-relation}
\tan(\beta+\tilde\alpha)\simeq\frac{\alpha_b}{\alpha_c}.
\end{equation}
Thus, $\alpha_b\alpha_c>0$ when
\begin{equation}
0<\beta+\tilde\alpha<\half\pi,\qquad\text{First (or third) quadrant}
\end{equation}
whereas $\alpha_b\alpha_c<0$ when
\begin{equation}
-\half\pi<\beta+\tilde\alpha<0\quad\text{or}\quad 
\half\pi<\beta+\tilde\alpha<\pi,\qquad\text{Second (or fourth) quadrant}
\end{equation}
in agreement with fig.~\ref{Fig:alphas-1}.  Finally, we note from
(\ref{Eq:approx-angular-relation}) that the physical region may occupy a large
fraction of the actual quadrant when $|\beta+\tilde\alpha|\simeq\pi/2$,
and only a small fraction of the quadrant when 
$|\beta+\tilde\alpha|\simeq0$ or $\pi$.
%%%%%%%%%%%%%%%%%%%%%%%%%%%%%%%%%%%%%%%%%%%%%%%%%%%%%%%%%%%%%%%%%%%%%%%%%%%%%%
\subsubsection*{Higgs--vector--vector couplings}
%%%%%%%%%%%%%%%%%%%%%%%%%%%%%%%%%%%%%%%%%%%%%%%%%%%%%%%%%%%%%%%%%%%%%%%%%%%%%%
The non-discovery of a Higgs boson at LEP poses constraints on $\tan\beta$ and
on the Higgs mass. While SM Higgs masses below 115~GeV are excluded
\cite{Heister:2001kr}, as well as $\tan\beta\lsim2$--3 in the MSSM 
\cite{Achard:2002zr}, these bounds
can be eluded in the 2HDM, if the lightest Higgs boson couples with
sufficiently reduced strength to the vector bosons. In fact, this coupling is
(relative to the corresponding SM coupling) given by
\begin{equation} \label{Eq:HZZ-couplings}
H_i ZZ: \qquad \cos\beta\, R_{i1} +\sin\beta\, R_{i2}.
\end{equation}
Higgs masses down to about 50~GeV are allowed, provided
this coupling squared is less than about 0.5 \cite{Abbiendi:2002qp}.
%%%%%%%%%%%%%%%%%%%%%%%%%%%%%%%%%%%%%%%%%%%%%%%%%%%%%%%%%%%%%%%%%%%%%%%%
\begin{figure}[htb]
\refstepcounter{figure}
\label{Fig:gvv_sq-2hdm}
\addtocounter{figure}{-1}
\begin{center}
\setlength{\unitlength}{1cm}
\begin{picture}(16.0,5.0)
\put(-1,0)
{\mbox{\epsfysize=5.0cm
\epsffile{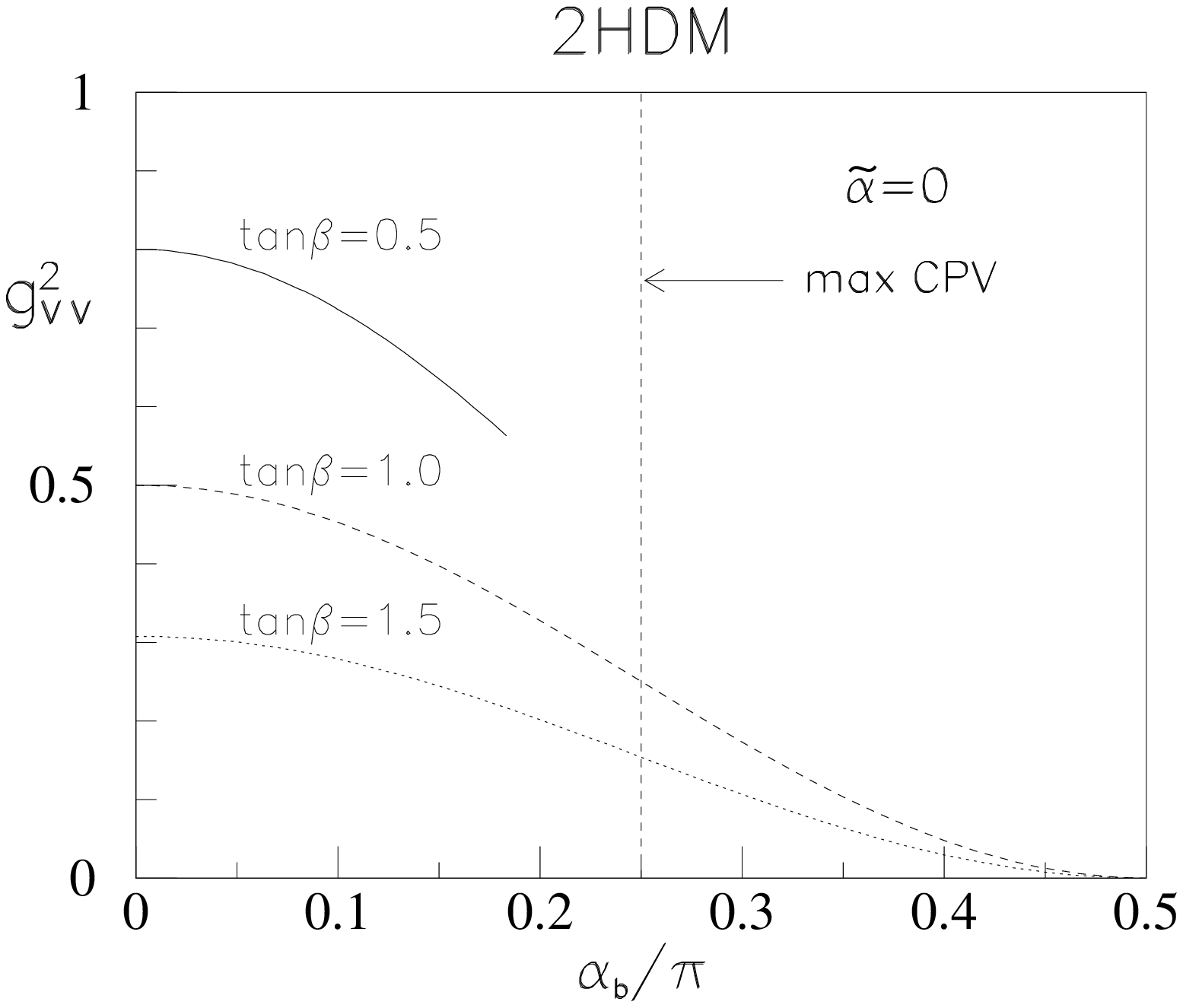}}
 \mbox{\epsfysize=5.0cm
\epsffile{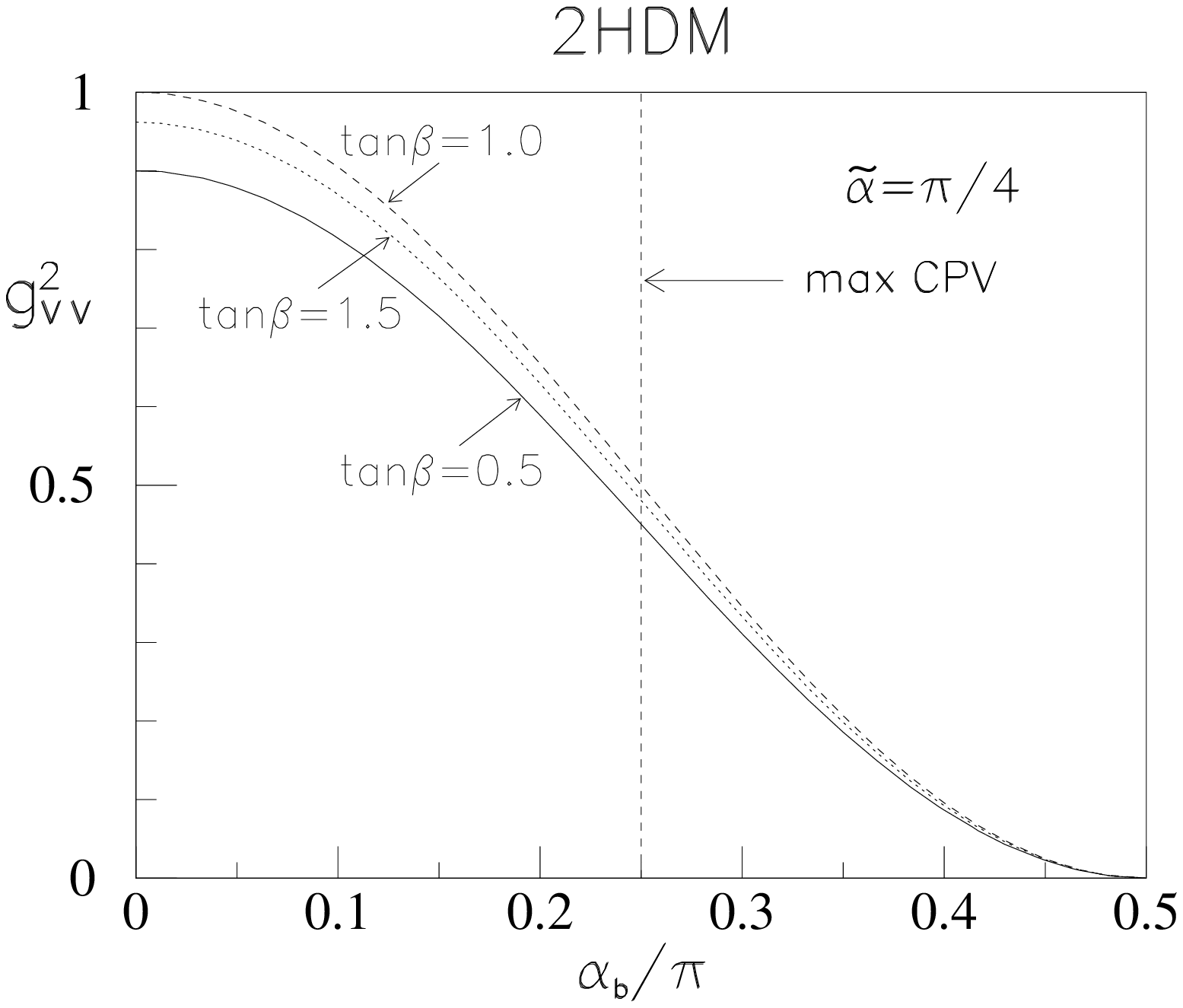}}
\mbox{\epsfysize=5.0cm
\epsffile{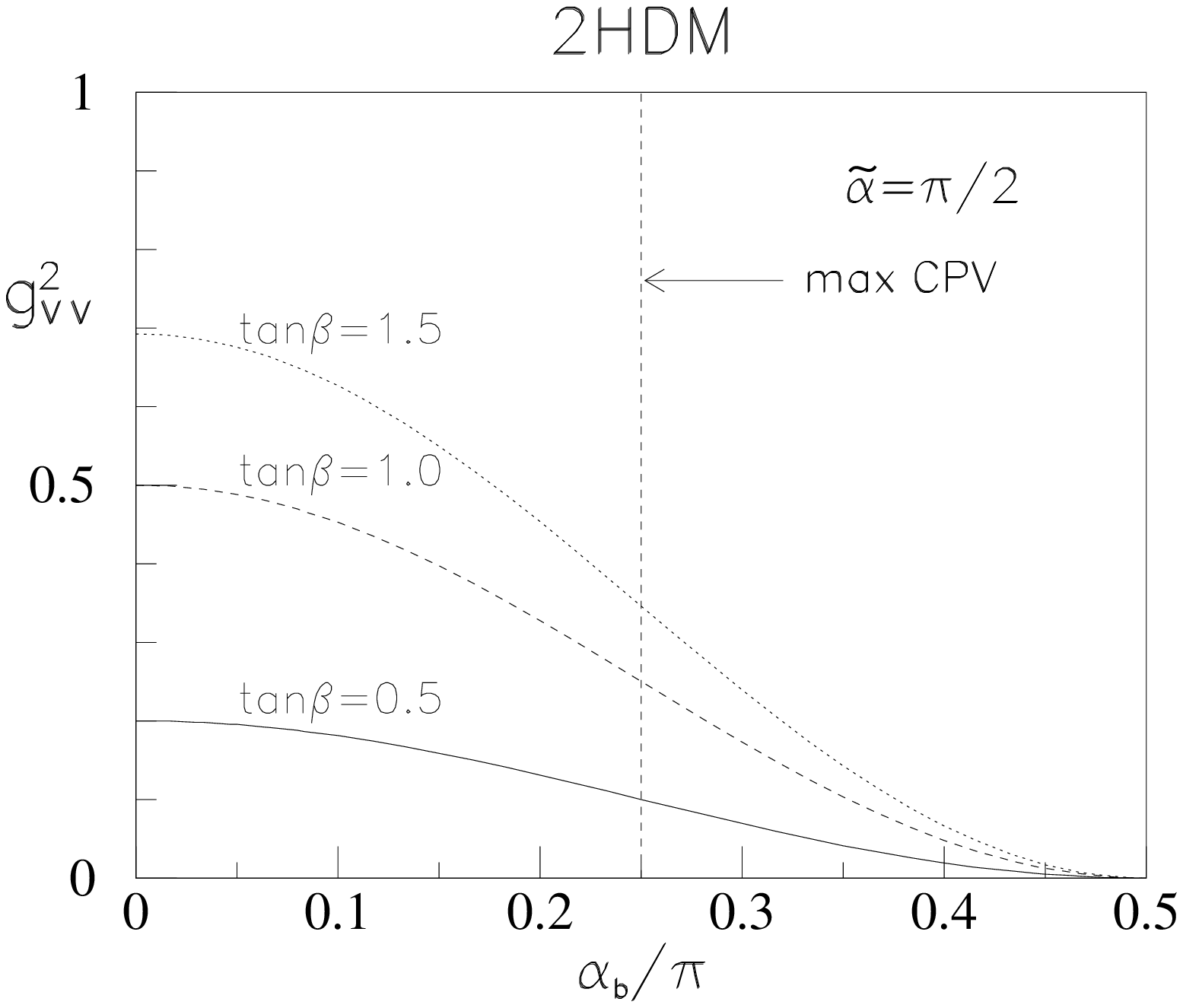}}}
\end{picture}
%\vspace*{-4mm}
\caption{Squared coupling of the {\it lightest} Higgs boson to vector bosons
(relative to that of the Standard Model),
$g_{VV}^2$, {\it vs.}\ mixing angle $\alpha_b$, for fixed $\tilde\alpha=0$,
$\pi/4$ and $\pi/2$.
These couplings are symmetric under $\alpha_b\to\pi-\alpha_b$,
and independent of $\alpha_c$. For $\tilde\alpha=0$ and $\tan\beta=0.5$,
only small values of $\alpha_b$ are allowed for the chosen mass parameters:
$M_1=100~\text{GeV}$, $M_2=300~\text{GeV}$, $M_{H^\pm}=500~\text{GeV}$, 
$\alpha_c=\pi/3$ and $\xi=0.8$ (cf.\ fig.~\ref{Fig:alphas-1}).}
\end{center}
\end{figure}
%%%%%%%%%%%%%%%%%%%%%%%%%%%%%%%%%%%%%%%%%%%%%%%%%%%%%%%%%%%%%%%%%%%%%%

We show in fig.~\ref{Fig:gvv_sq-2hdm} the square of the coupling of the
lightest Higgs boson to vector bosons, relative to that of the Standard Model,
$g_{VV}^2$, for low values of $\tan\beta$. As discussed above, for masses of
the lightest Higgs below $\sim100~\text{GeV}$, this quantity should be well
below unity, in order not to be in conflict with the non-observation of a
Higgs boson at LEP2 \cite{Heister:2001kr}.

It should be noted that this coupling can be expressed in terms of angles
only, without any dependence on masses. However, for certain combinations of
mass parameters and $\alpha_c$ values, some $\alpha_b$ ranges might be
unphysical, as illustrated for $\tilde\alpha=0$ and $\tan\beta=0.5$ in this
figure, as well as in fig.~\ref{Fig:alphas-1}.

%%%%%%%%%%%%%%%%%%%%%%%%%%%%%%%%%%%%%%%%%%%%%%%%%%%%%%%%%%%%%%%%%%%%%%%%%%%%%%
\section{Case Studies}
\setcounter{equation}{0}
%%%%%%%%%%%%%%%%%%%%%%%%%%%%%%%%%%%%%%%%%%%%%%%%%%%%%%%%%%%%%%%%%%%%%%%%%%%%%%
In order to get a better idea how much $CP$ violation the 2HDM can give,
we shall here discuss the case of ``maximal $CP$ violation'' in the sense
of eq.~(\ref{Eq:max_CP}), together with small values of $\tan\beta$ and
light Higgs masses.

In the following two subsections, we shall analyze two cases, first
(sec.~\ref{Sec:higgs-light-light-heavy}) the case of two Higgs bosons being
light, and then (sec.~\ref{Sec:higgs-light-heavy-heavy}) the case of one light
and two heavy ones. These are qualitatively different, since in the first case
there will be considerable cancellations among the $CP$-violating
contributions, because of (\ref{Eq:orthogonal}).

With all mixing angles fixed, and two Higgs masses specified, the third Higgs
mass is determined. Thus, the ``soft'' mass term $\mu^2=\Re
m_{12}^2/\sin2\beta$ only enters in converting the masses (and angles) to
$\lambda$'s. We check that the required $\lambda$'s satisfy positivity of the
potential and perturbativity, eq.~(\ref{Eq:pert-xi}).

%%%%%%%%%%%%%%%%%%%%%%%%%%%%%%%%%%%%%%%%%%%%%%%%%%%%%%%%%%%%%%%%%%%%%%%%%%%%%%
\subsection{Case 1. Two light and one heavy Higgs bosons} 
\label{Sec:higgs-light-light-heavy}
%%%%%%%%%%%%%%%%%%%%%%%%%%%%%%%%%%%%%%%%%%%%%%%%%%%%%%%%%%%%%%%%%%%%%%%%%%%%%%
As a first case, we consider $M_1\le M_2=300~\text{GeV}$ (below the $t\bar t$
threshold) with $M_{H^\pm}=500~\text{GeV}$ and $\tan\beta=0.5$ and 1.0.  (This
is well within the limits on $M_{H^\pm}$ and $\tan\beta$ derived from studies
of meson decays and mixings \cite{Hewett:1996ct}.)  Then, for
$\alpha_c=\pi/4$, $\pi/3$ and $5\pi/12$, we show in
fig.~\ref{Fig:2hdm-masses-300} the heaviest Higgs boson mass, $M_3$ {\it vs.}\
$M_1$, for $\tan\beta=0.5$, 1.0 and 1.5.  We note that, for some of the
parameters considered, there is no solution (we take $\xi=1.0$), as also
follows from fig.~\ref{Fig:alphas-1}.
%%%%%%%%%%%%%%%%%%%%%%%%%%%%%%%%%%%%%%%%%%%%%%%%%%%%%%%%%%%%%%%%%%%%%%%%
\begin{figure}[htb]
\refstepcounter{figure}
\label{Fig:2hdm-masses-300}
\addtocounter{figure}{-1}
\begin{center}
\setlength{\unitlength}{1cm}
\begin{picture}(10.0,7.0)
\put(0,0)
{\mbox{\epsfysize=7.5cm
\epsffile{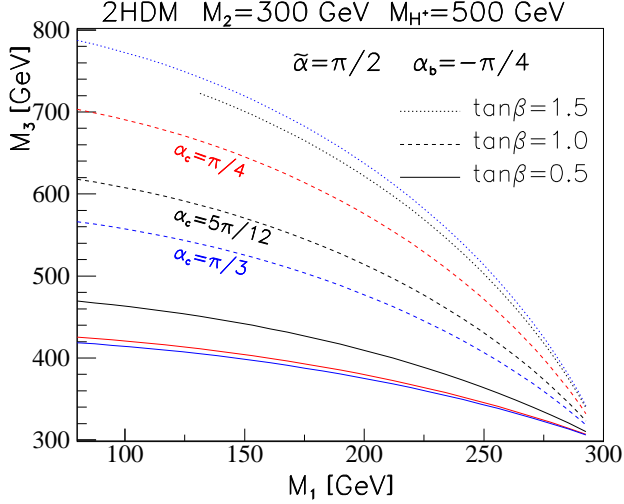}}}
\end{picture}
%\vspace*{-4mm}
\caption{Heaviest Higgs-boson mass, $M_3$ {\it vs.}\ $M_1$, for
$M_2=300~\text{GeV}$, and $M_{H^\pm}=500~\text{GeV}$, for three values of
$\alpha_c=\pi/4$, $\pi/3$ and $5\pi/12$, and three values of
$\tan\beta=0.5$, 1.0 and 1.5.  Furthermore, we consider
``maximal $CP$ violation'' with $\tilde\alpha=\pi/2$ and $\alpha_b=-\pi/4$.}
\end{center}
\end{figure}
%%%%%%%%%%%%%%%%%%%%%%%%%%%%%%%%%%%%%%%%%%%%%%%%%%%%%%%%%%%%%%%%%%%%%%

In fig.~\ref{Fig:a1-2hdm-300-0.5} we show, for this case of ``maximal $CP$
violation'' and low values of $\tan\beta$ ($\tan\beta=0.5$ and 1.0), the
corresponding signal-to-noise ratio of the observable $A_1$ (\ref{Eq:A_1}),
for $M_1$ ranging from 80 to 300~GeV, keeping $M_2=300~\text{GeV}$ fixed and
letting $M_3$ vary accordingly. We show separately the contributions of only
the lightest Higgs boson, and the contributions of all three.  Considering
only the contribution of the lightest Higgs boson, the $CP$-violating effects
are seen to be significantly larger than in the ``general case'' studied in
sect.~\ref{sect:model-indep}, essentially because of the enhancement of
$\gamma_{CP}$ due to the small values of $\tan\beta$.  We note that, for the
parameters considered here ($\tilde\alpha=\pi/2$ and $\alpha_b=-\pi/4$), and
for the lightest Higgs boson, $\gamma_{CP}$ is negative, contrary to the case
studied in sect.~\ref{sect:model-indep}.
%%%%%%%%%%%%%%%%%%%%%%%%%%%%%%%%%%%%%%%%%%%%%%%%%%%%%%%%%%%%%%%%%%%%%%%%
\begin{figure}[htb]
\refstepcounter{figure}
\label{Fig:a1-2hdm-300-0.5}
\addtocounter{figure}{-1}
\begin{center}
\setlength{\unitlength}{1cm}
\begin{picture}(10.0,7.0)
\put(-4.0,0)
{\mbox{\epsfysize=7.5cm
\epsffile{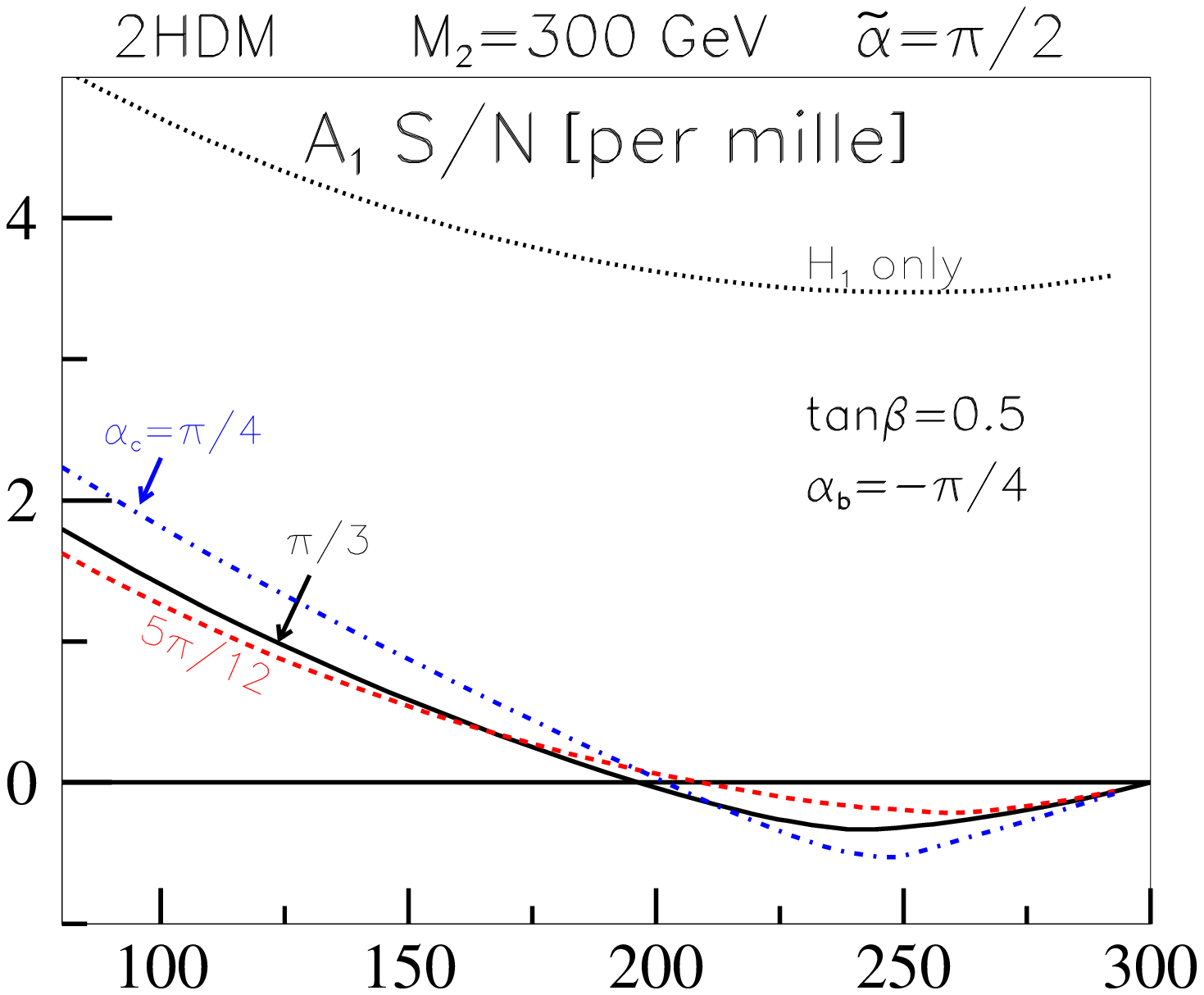}}
 \mbox{\epsfysize=7.5cm
\epsffile{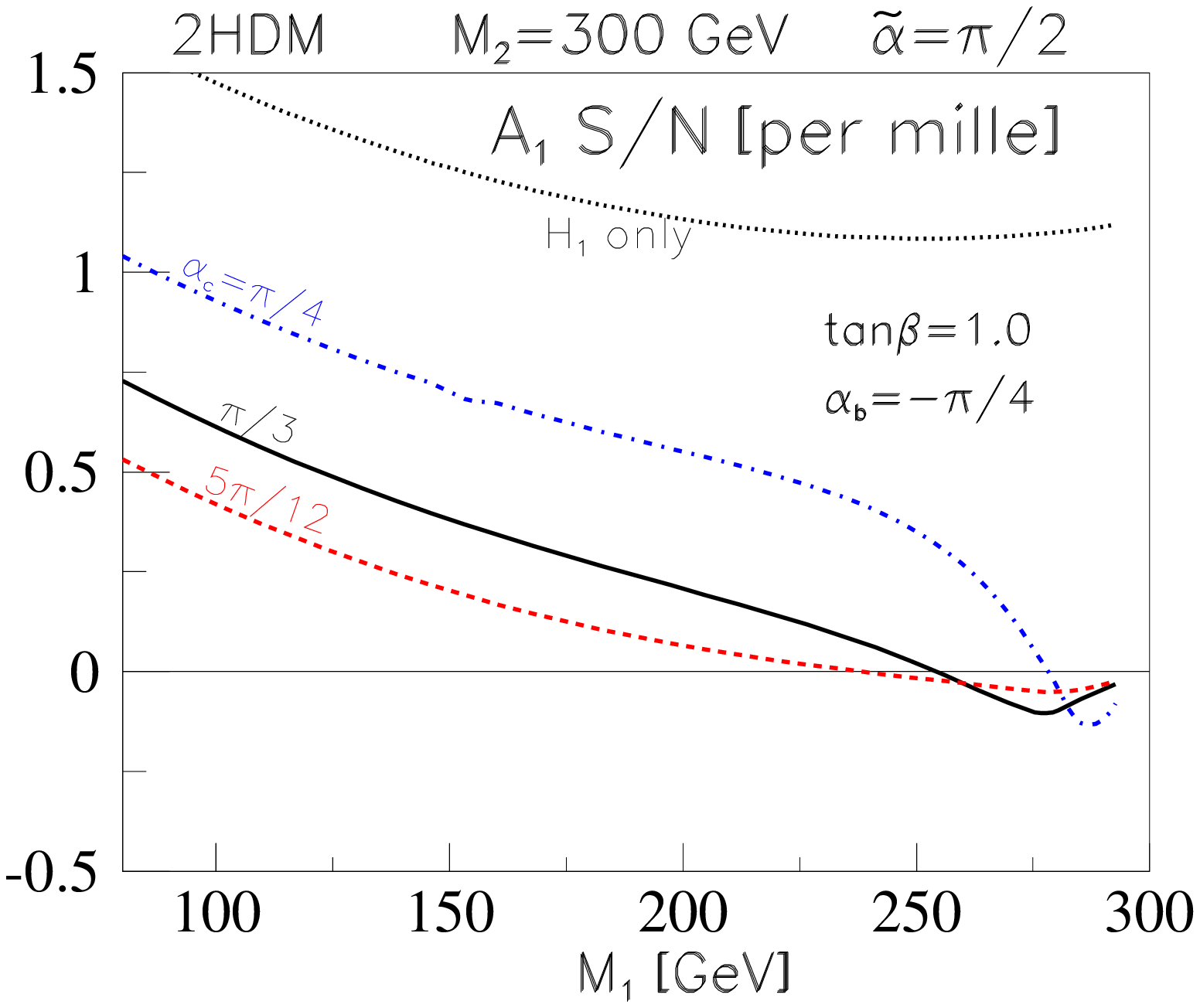}}}
\end{picture}
%\vspace*{-4mm}
\caption{Signal-to-noise ratio, for lepton energy correlations $\langle
A_1\rangle$ in $pp\to t\bar tX$ {\it vs.}\ lightest Higgs mass for the 2HDM,
for a case of three moderately light Higgs bosons (see
fig.~\ref{Fig:2hdm-masses-300}) and $\tan\beta=0.5$ and 1.0.  ``Maximal $CP$
violation'' is considered (cf.\ eq.~(\ref{Eq:max_CP})), and $\alpha_c=\pi/4$,
$\pi/3$ and $5\pi/12$. Also shown, is the contribution of the lightest Higgs
boson only (independent of $\alpha_c$).}
\end{center}
\end{figure}
%%%%%%%%%%%%%%%%%%%%%%%%%%%%%%%%%%%%%%%%%%%%%%%%%%%%%%%%%%%%%%%%%%%%%%%%

However, in the present case, the inclusion of all three contributions leads
to large cancellations.  This also implies that, even if no $CP$ violation
should be observed, it may be difficult to conclude that the Higgs sector
conserves $CP$, since such cancellations are possible.

As expected, the $CP$-violating effects are largest for low values of $M_1$.
Since the two lightest Higgs bosons are below the $t\bar t$ resonance, there
is a smooth dependence on $M_1$.  As $M_1$ approaches $M_2$ (where all three
masses are degenerate), the $CP$-violating effects cancel.  On the whole, the
resulting $CP$ violation is reduced by a factor of $\sim1/4$ compared to the
``model-independent'' case considered in sect.~3.

Whereas the top Yukawa couplings of the {\it lightest} Higgs boson do not
depend on $\alpha_c$, some dependence on this quantity enters, via the
couplings of the two heavier Higgs bosons to the $t$ quark.  This is also
illustrated in fig.~\ref{Fig:a1-2hdm-300-0.5}.
%%%%%%%%%%%%%%%%%%%%%%%%%%%%%%%%%%%%%%%%%%%%%%%%%%%%%%%%%%%%%%%%%%%%%%%%
\begin{figure}[htb]
\refstepcounter{figure}
\label{Fig:a1-2hdm-300-al_til}
\addtocounter{figure}{-1}
\begin{center}
\setlength{\unitlength}{1cm}
\begin{picture}(10.0,7.0)
\put(-3.5,0)
{\mbox{\epsfysize=7.5cm
\epsffile{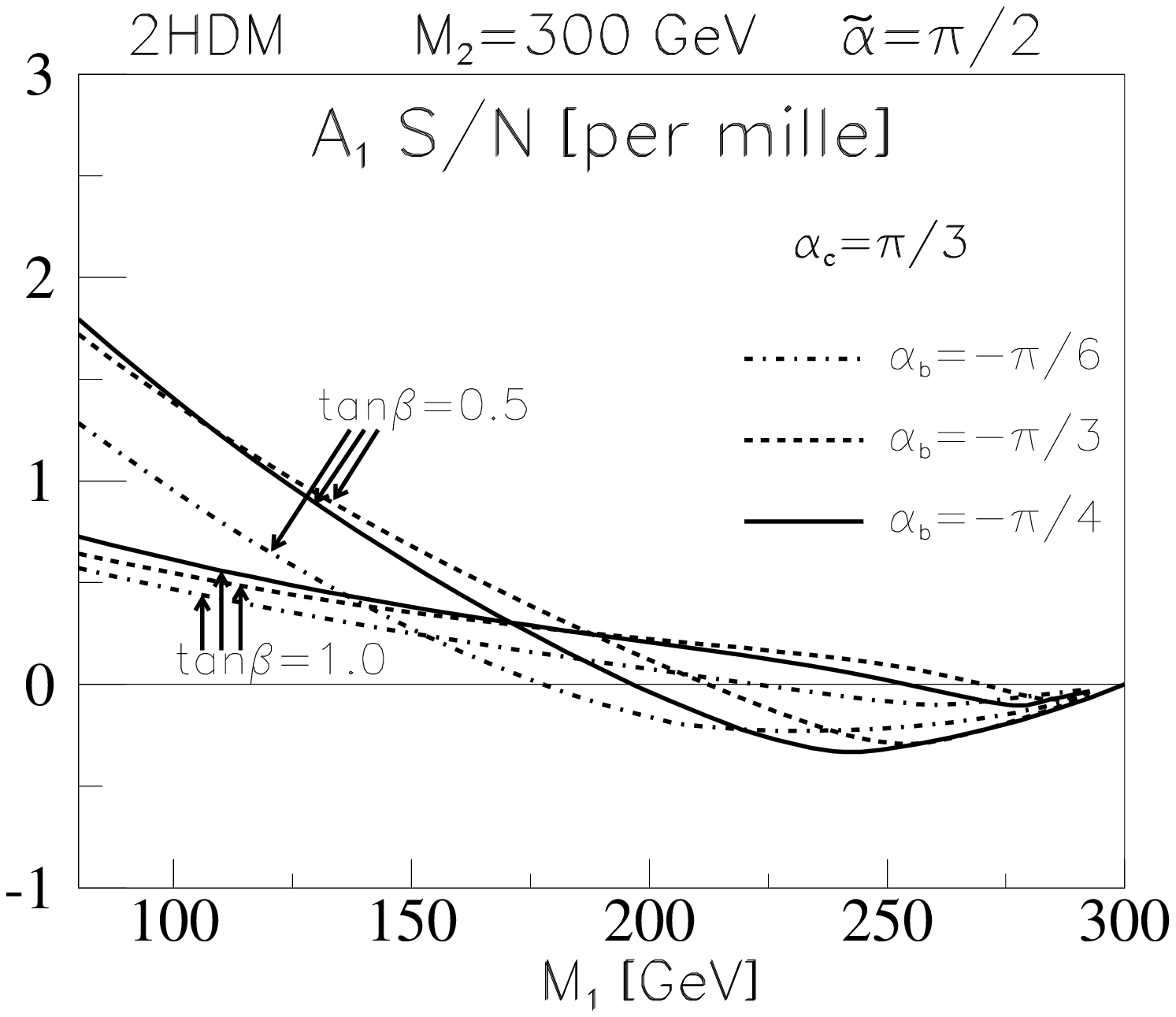}}
 \mbox{\epsfysize=7.5cm
\epsffile{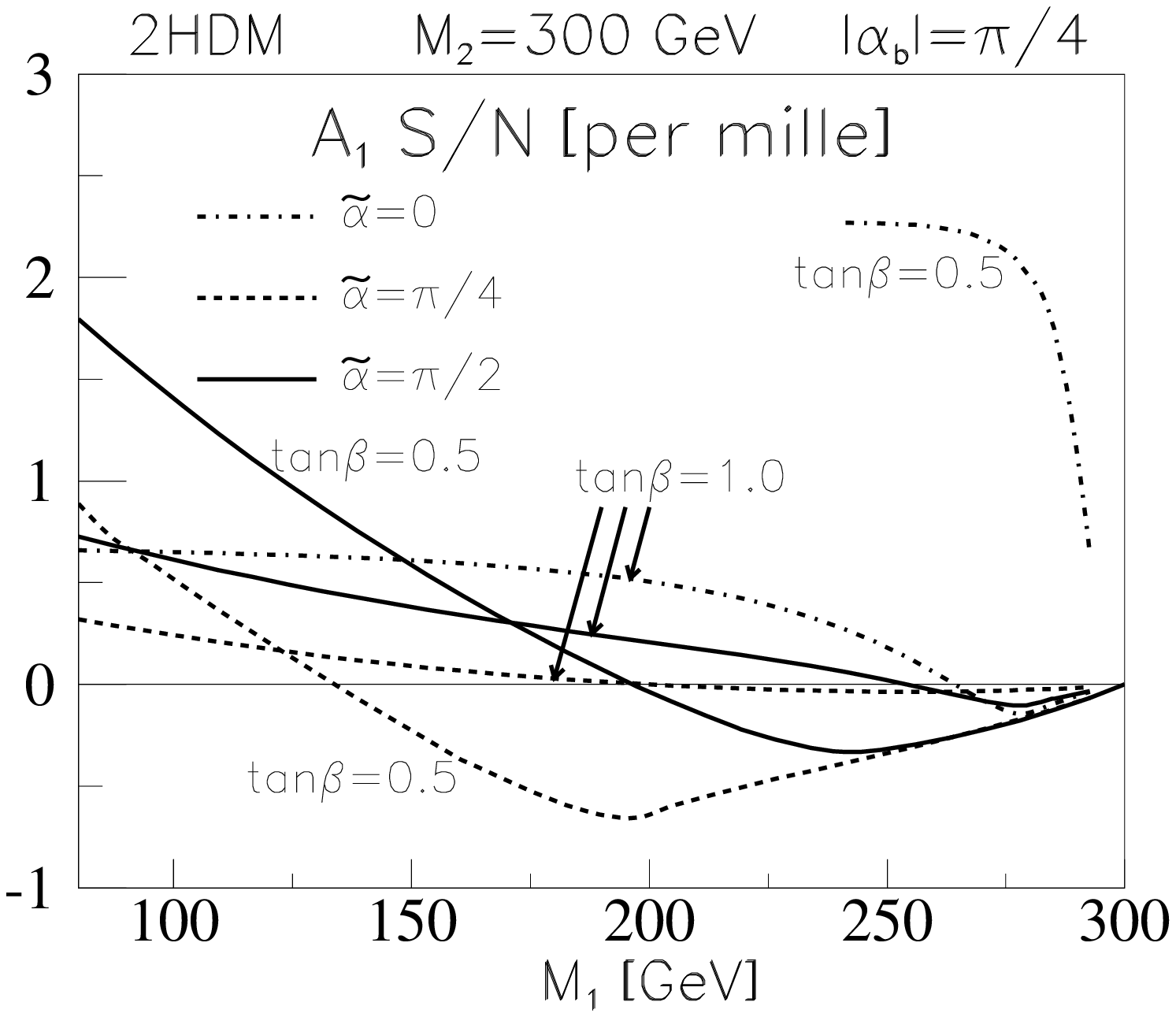}}}
\end{picture}
%\vspace*{-4mm}
\caption{Sensitivity of $A_1$ in $pp\to t\bar tX$ {\it vs.}\ lightest Higgs
mass.  {\it Left panel:} Different values of $\alpha_b$ are studied, for fixed
$\tilde\alpha=\pi/2$, $\alpha_c=\pi/3$.  {\it Right panel:} Different values
of $\tilde\alpha$ are studied, $\tilde\alpha=0$, $\pi/4$ and $\pi/2$
(``maximal mixing'') with $\alpha_c=\pi/3$ (see fig.~\ref{Fig:alphas-1}).  The
mixing angle $|\alpha_b|=\pi/4$ is kept fixed. Two values of $\tan\beta$ are
considered: 0.5 and 1.0.}
\end{center}
\end{figure}
%%%%%%%%%%%%%%%%%%%%%%%%%%%%%%%%%%%%%%%%%%%%%%%%%%%%%%%%%%%%%%%%%%%%%%

Since the two heavier Higgs bosons can contribute significantly to
$\langle A_1\rangle$, the dependence on $\alpha_b$ need not be as simple as
discussed in sect.~\ref{sect-parametrization}. In particular, the $CP$
violation need not be maximal for $\alpha_b=\pm\pi/4$ (or $\pm3\pi/4$).
Fig.~\ref{Fig:a1-2hdm-300-al_til} (left panel) illustrates the dependence on
$\alpha_b$ for fixed $\tilde\alpha=\pi/2$, $\alpha_c=\pi/3$, and two values of
$\tan\beta$ (0.5 and 1.0).  One confirms the general impression from the form
of (\ref{Eq:R-angles}) and the discussion following (\ref{Eq:max-mix}), that
$|\alpha_b|$ should be sizable, but not necessarily $\pi/4$, in order to
maximize the $CP$ violation.

It is also possible that one Higgs boson does not violate $CP$ in its
couplings to the $t$ quark, and yet the other two do.  For example, for
$\tilde\alpha=0$, the lightest Higgs boson has a pure pseudoscalar coupling to
the top quark, and hence the $CP$ violation is exclusively due to the two
heavier Higgs bosons. In this case, the resulting total contribution to the
lepton energy correlations $\langle A_1\rangle$ may even exceed that obtained
when the contribution of the lightest Higgs boson is maximal, see
fig.~\ref{Fig:a1-2hdm-300-al_til} (right panel).
%%%%%%%%%%%%%%%%%%%%%%%%%%%%%%%%%%%%%%%%%%%%%%%%%%%%%%%%%%%%%%%%%%%%%%%%%%%%%%
\subsection{Case 2. One light and two heavy neutral Higgs bosons}
\label{Sec:higgs-light-heavy-heavy}
%%%%%%%%%%%%%%%%%%%%%%%%%%%%%%%%%%%%%%%%%%%%%%%%%%%%%%%%%%%%%%%%%%%%%%%%%%%%%%
Next, we let two of the neutral Higgs boson masses be more heavy, taking
$M_3\ge M_2=500~\text{GeV}$, and $M_{H^\pm}=700~\text{GeV}$.  The resulting
mass values, $M_3$, are shown in fig.~\ref{Fig:2hdm-masses-500}, for two
values of $\tilde\alpha$, three values of $\alpha_c=\pi/4$, $\pi/3$ and
$5\pi/12$, and small values of $\tan\beta$.  For the cases studied in
sect.~\ref{Sec:higgs-light-light-heavy}, solutions could be found with
$\mu=0$. This is here not the case. For $\tilde\alpha=\pi/2$ (``maximal $CP$
violation''), {\it no} value of $\mu$ allows a light $H_1$
(fig.~\ref{Fig:2hdm-masses-500}, right panel), so we have included also
$\tilde\alpha=\pi/3$ (left panel), where light $H_1$ can be
realized for large $\mu$.
%%%%%%%%%%%%%%%%%%%%%%%%%%%%%%%%%%%%%%%%%%%%%%%%%%%%%%%%%%%%%%%%%%%%%%%%
\begin{figure}[htb]
\refstepcounter{figure}
\label{Fig:2hdm-masses-500}
\addtocounter{figure}{-1}
\begin{center}
\setlength{\unitlength}{1cm}
\begin{picture}(10.0,7.0)
\put(-3.8,0)
{\mbox{\epsfysize=7.5cm
\epsffile{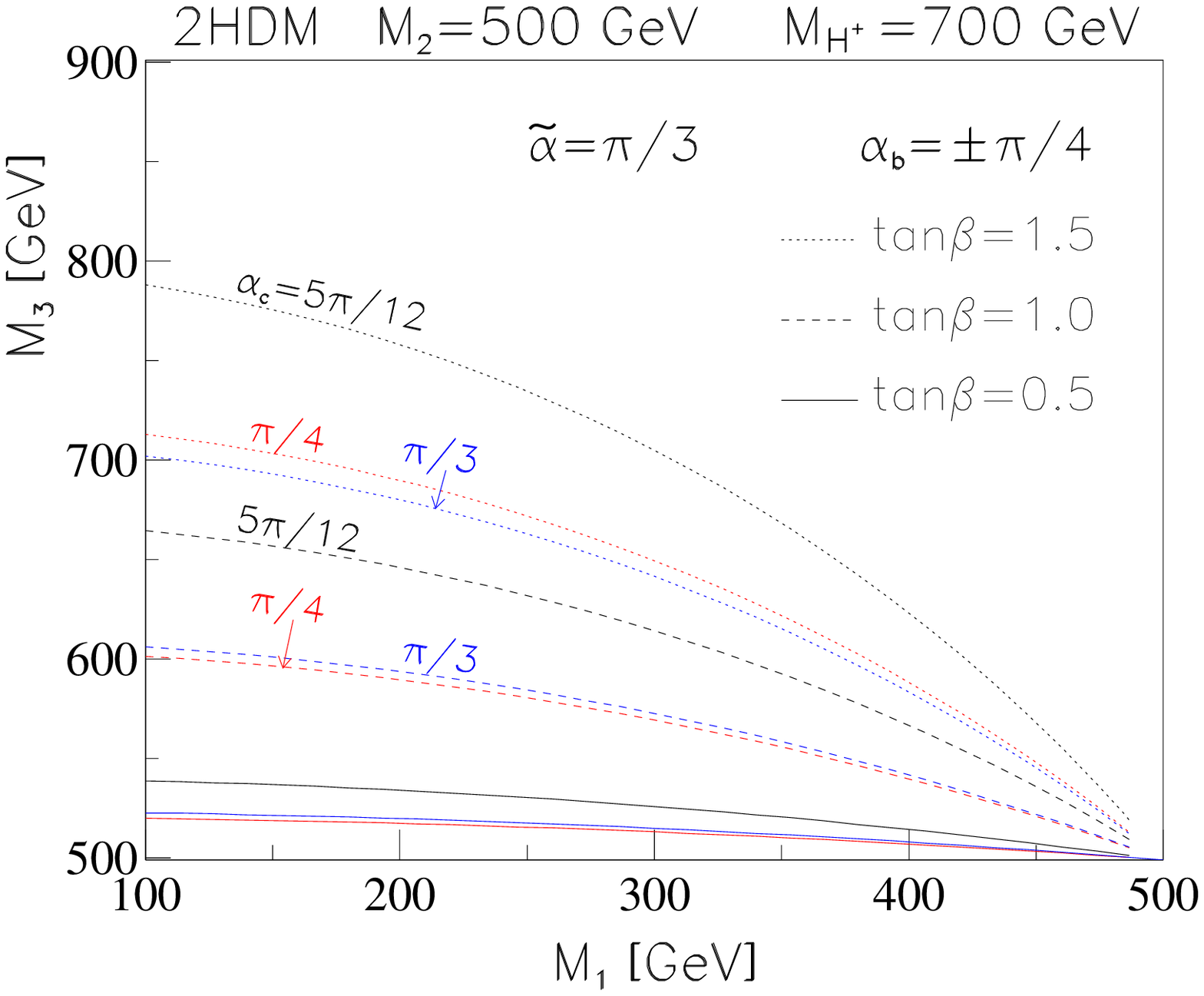}}
 \mbox{\epsfysize=7.5cm
\epsffile{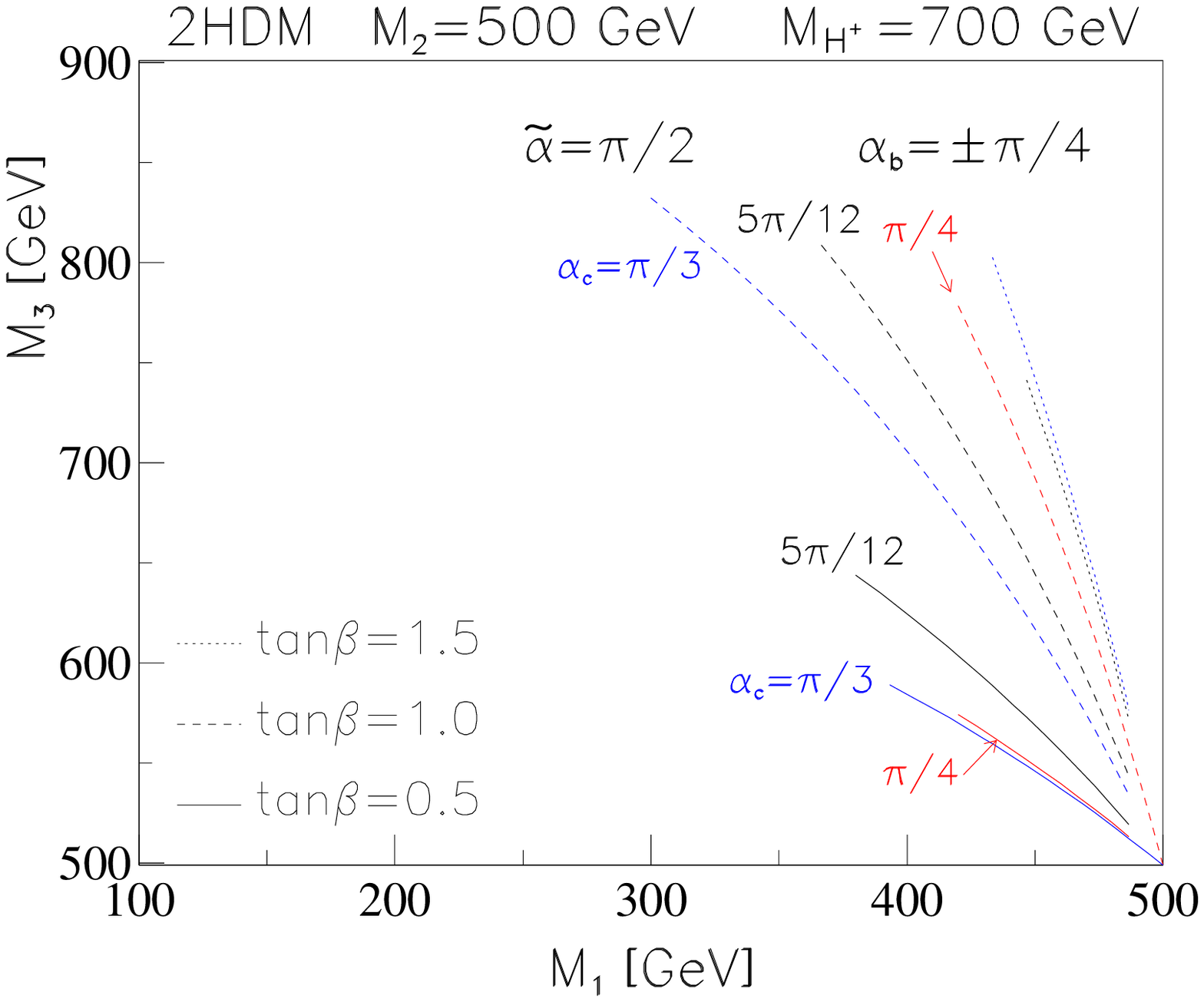}}}
\end{picture}
%\vspace*{-4mm}
\caption{Heaviest Higgs-boson mass, $M_3$ {\it vs.}\ $M_1$, for
$M_2=500~\text{GeV}$ and $M_{H^\pm}=700~\text{GeV}$.
For some parameters, there is a lower limit for $M_1$, due
to perturbativity, eq.~(\ref{Eq:pert-xi}), where we take $\xi=1.0$.}
\end{center}
\end{figure}
%%%%%%%%%%%%%%%%%%%%%%%%%%%%%%%%%%%%%%%%%%%%%%%%%%%%%%%%%%%%%%%%%%%%%%

We show in fig.~\ref{Fig:a1-2hdm-500-0.5} the corresponding results for the
lepton energy asymmetry, $\langle A_1\rangle$, for $\tan\beta=0.5$ and 1.  In
this case, as opposed to the cases considered in
sect.~\ref{Sec:higgs-light-light-heavy}, since $H_2$ and $H_3$ are heavier, it
is a good approximation to consider only $H_1$ exchange up to $M_1\sim
450~\text{GeV}$.  As a result, the $CP$-violating effects are significantly
larger than in the low-mass case studied in
sect.~\ref{Sec:higgs-light-light-heavy}.  The amount of $CP$-violation for {\it
one} light Higgs boson, $M_1={\cal O}(100~\text{GeV})$, is comparable with,
and may exceed that obtained for a mass around the $t\bar t$ resonance.
%%%%%%%%%%%%%%%%%%%%%%%%%%%%%%%%%%%%%%%%%%%%%%%%%%%%%%%%%%%%%%%%%%%%%%%%
\begin{figure}[htb]
\refstepcounter{figure}
\label{Fig:a1-2hdm-500-0.5}
\addtocounter{figure}{-1}
\begin{center}
\setlength{\unitlength}{1cm}
\begin{picture}(10.0,7.0)
\put(-3.5,0)
{\mbox{\epsfysize=7.5cm
\epsffile{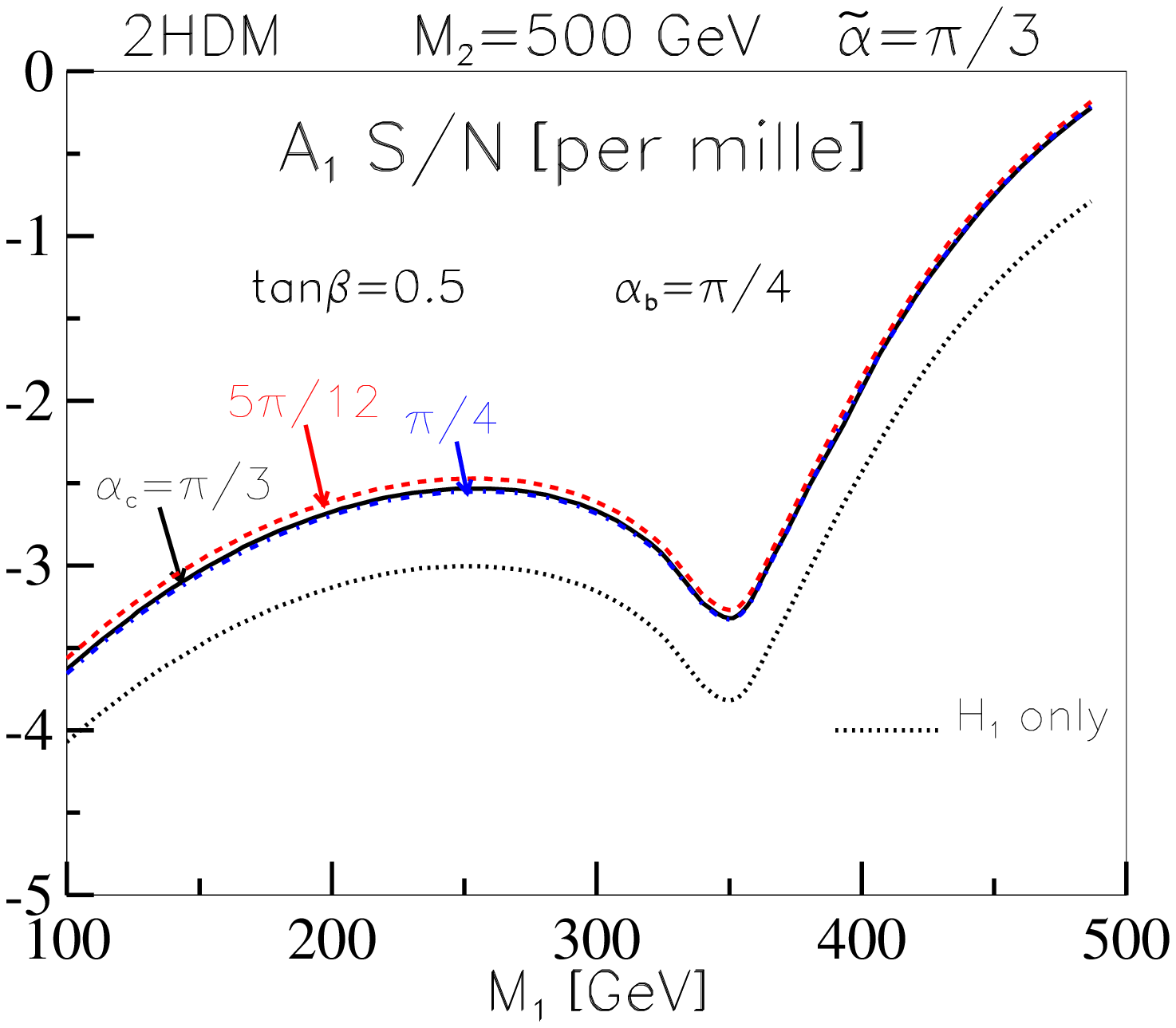}}
 \mbox{\epsfysize=7.5cm
\epsffile{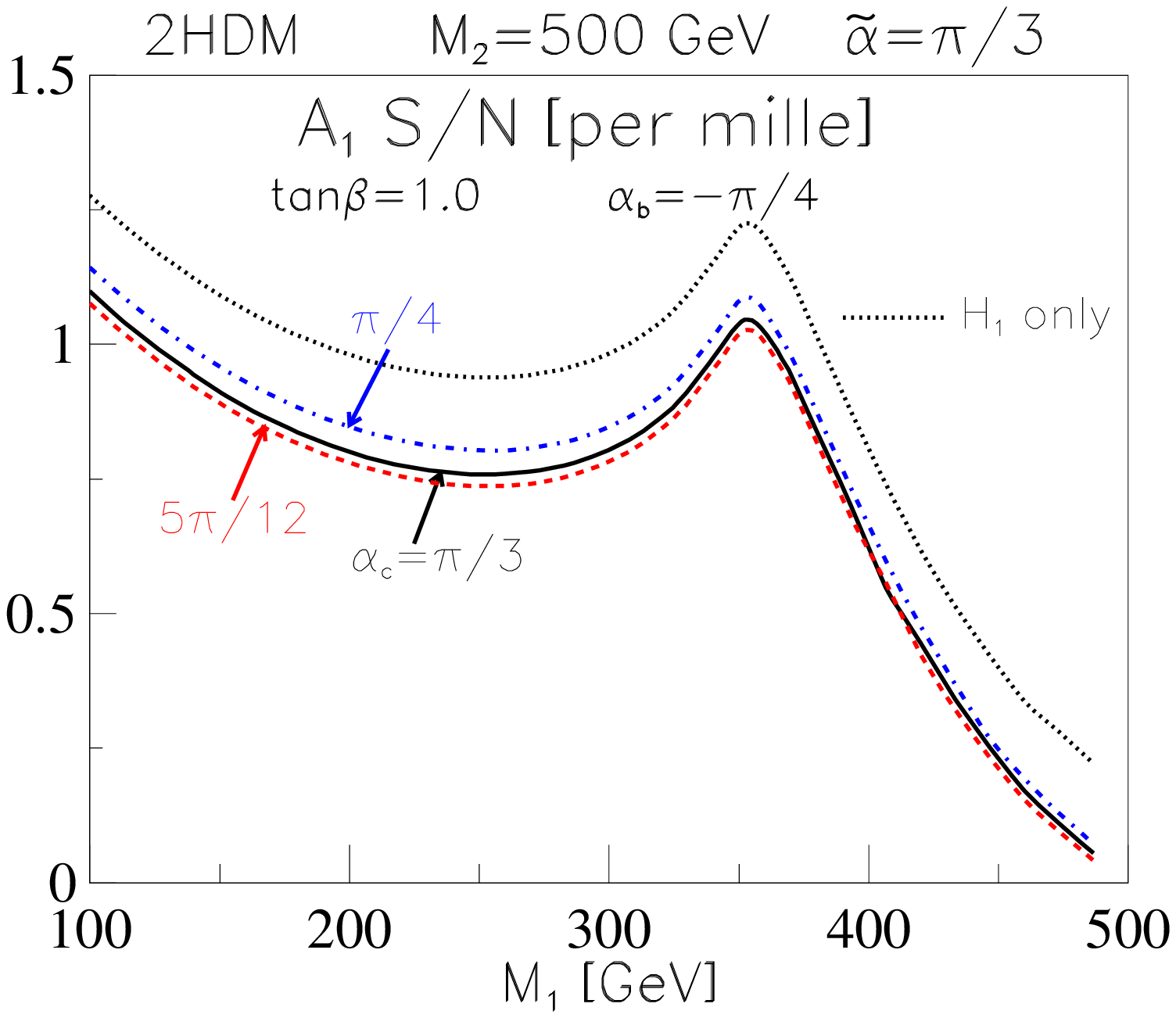}}}
\end{picture}
%\vspace*{-4mm}
\caption{Signal-to-noise ratio, for lepton energy correlations $\langle
A_1\rangle$ in $pp\to t\bar tX$ {\it vs.}\ Higgs mass for the 2HDM.  Similar
to fig.~\ref{Fig:a1-2hdm-300-0.5} for $\tilde\alpha=\pi/3$ and
$M_3>M_2=500~\text{GeV}$, with $M_{H^\pm}=700~\text{GeV}$.
Note that one set of curves is ``upside--down'' with respect to the other,
since the $\alpha_b$-values have opposite signs
(we keep $\alpha_c$ positive in both cases).}
\end{center}
\end{figure}
%%%%%%%%%%%%%%%%%%%%%%%%%%%%%%%%%%%%%%%%%%%%%%%%%%%%%%%%%%%%%%%%%%%%%%

We note that there is only a rather weak dependence on $\alpha_c$.
This is clearly due to the fact that the contribution of $H_1$
(which is independent of $\alpha_c$) dominates.
(However, as $\alpha_c\to0$, the physical range of $M_1$ shrinks,
cf.\ fig.~\ref{Fig:alphas-1}.)

%%%%%%%%%%%%%%%%%%%%%%%%%%%%%%%%%%%%%%%%%%%%%%%%%%%%%%%%%%%%%%%%%%%%%%%%
\begin{figure}[htb]
\refstepcounter{figure}
\label{Fig:a1-2hdm-500-al_b-0.5}
\addtocounter{figure}{-1}
\begin{center}
\setlength{\unitlength}{1cm}
\begin{picture}(10.0,7.0)
\put(0,0)
{\mbox{\epsfysize=7.5cm
\epsffile{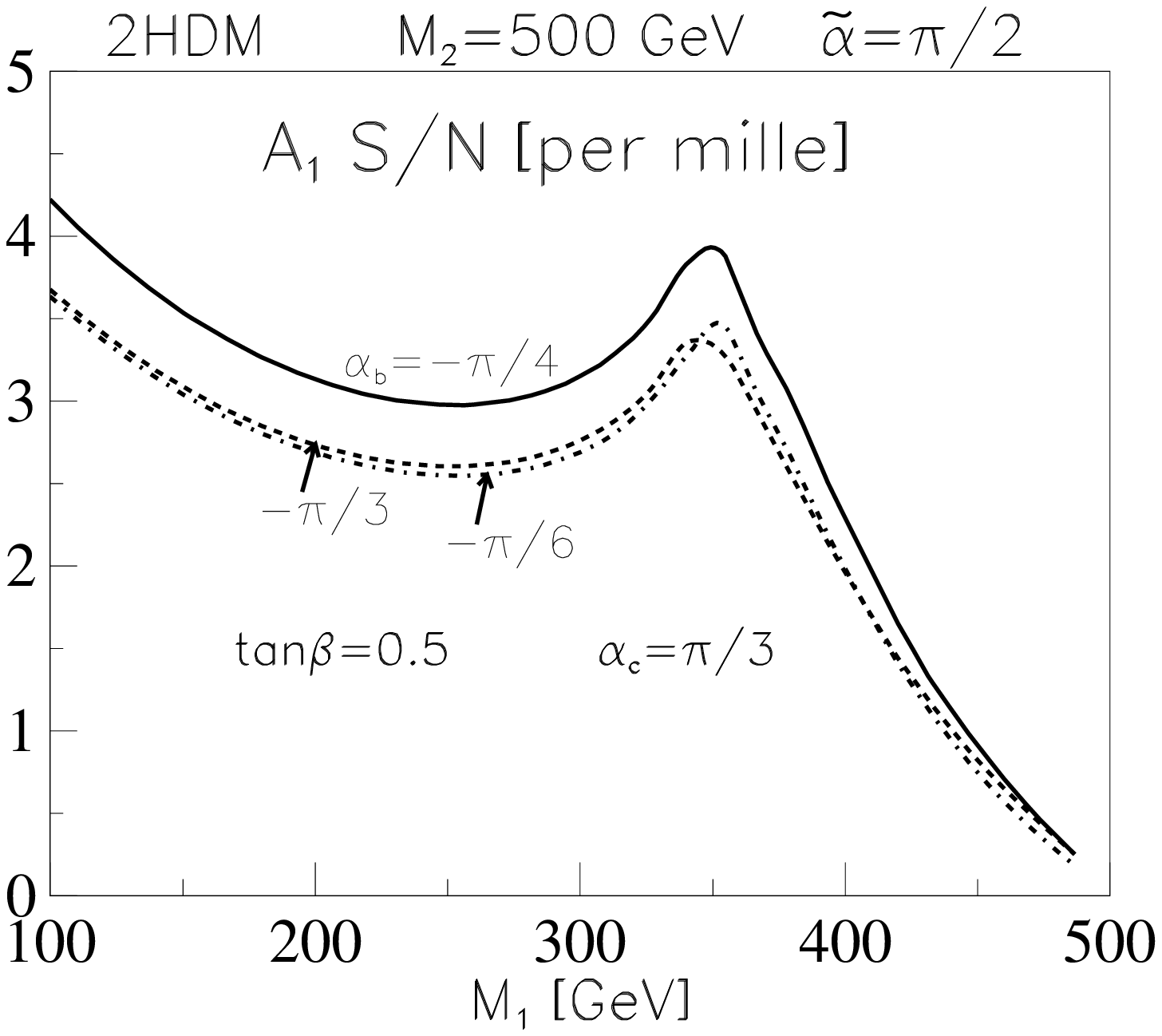}}}
\end{picture}
%\vspace*{-4mm}
\caption{Signal-to-noise ratio, for lepton energy correlations $\langle
A_1\rangle$ in $pp\to t\bar tX$ {\it vs.}\ Higgs mass for the 2HDM. Similar to
fig.~\ref{Fig:a1-2hdm-500-0.5} for three values of $\alpha_b$.}
\end{center}
\end{figure}
%%%%%%%%%%%%%%%%%%%%%%%%%%%%%%%%%%%%%%%%%%%%%%%%%%%%%%%%%%%%%%%%%%%%%%

Fig.~\ref{Fig:a1-2hdm-500-al_b-0.5} is devoted to a study of the dependence
of $\alpha_b$. As expected (since $H_1$ dominates), the $CP$ violation
is maximized for $\alpha_b\simeq-\pi/4$.
As $\alpha_b\to0$, the effect vanishes. There are in fact two reasons
for this. First, the contribution of $H_1$ to $\gamma_{CP}$ vanishes
linearly. Secondly, $M_2$ and $M_3$ become degenerate in this limit.
According to the discussion in sect.~\ref{Sec:physical-content},
all three masses must then become degenerate, and there is no $CP$ violation.

%%%%%%%%%%%%%%%%%%%%%%%%%%%%%%%%%%%%%%%%%%%%%%%%%%%%%%%%%%%%%%%%%%%%%%%%%%%%%%
\subsection{\boldmath Case 3. The $t\bar t$ transition region}
\label{Sec:higgs-ttbar}
%%%%%%%%%%%%%%%%%%%%%%%%%%%%%%%%%%%%%%%%%%%%%%%%%%%%%%%%%%%%%%%%%%%%%%%%%%%%%%
As is seen from fig.~\ref{Fig:bb-cc}, and also from a qualitative comparison
of Case~1 and Case~2, there is in general less $CP$-violation if two Higgs
masses are below the $t\bar t$ threshold than if two masses are above. We here
explore this transition region in a little more detail, by comparing in
fig.~\ref{Fig:a1-2hdm-threshold} a few values of {\it the intermediate Higgs
boson mass} $M_2$, below and above the $t\bar t$ threshold, keeping the mixing
angles fixed at $\tilde\alpha=\pi/2$, $\alpha_b=-\pi/4$, $\alpha_c=\pi/3$ and
$\tan\beta=0.5$.  One sees that, for values of $M_1$ well below $M_2$, there
is little dependence on $M_2$, unless $M_2$ is {\it above} the $t\bar t$
threshold, in which case the $CP$-violation (for fixed $M_1$) increases
rapidly with $M_2$.
%%%%%%%%%%%%%%%%%%%%%%%%%%%%%%%%%%%%%%%%%%%%%%%%%%%%%%%%%%%%%%%%%%%%%%%%
\begin{figure}[htb]
\refstepcounter{figure}
\label{Fig:a1-2hdm-threshold}
\addtocounter{figure}{-1}
\begin{center}
\setlength{\unitlength}{1cm}
\begin{picture}(10.0,7.0)
\put(0,0)
{\mbox{\epsfysize=7.5cm
\epsffile{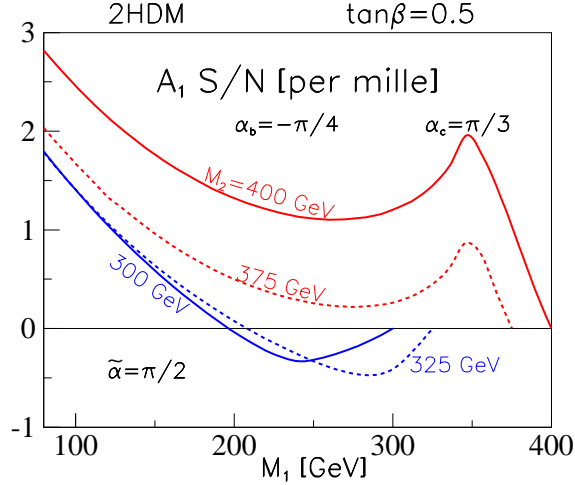}}}
\end{picture}
%\vspace*{-4mm}
\caption{Lepton energy correlations $\langle A_1\rangle$ in $pp\to t\bar tX$
{\it vs.}\ Higgs mass for the 2HDM. The intermediate Higgs mass,
$M_2$, is in the $t\bar t$ threshold region.}
\end{center}
\end{figure}
%%%%%%%%%%%%%%%%%%%%%%%%%%%%%%%%%%%%%%%%%%%%%%%%%%%%%%%%%%%%%%%%%%%%%%

%%%%%%%%%%%%%%%%%%%%%%%%%%%%%%%%%%%%%%%%%%%%%%%%%%%%%%%%%%%%%%%%%%%%%%%%%%%%%%
\section{Concluding remarks}
%%%%%%%%%%%%%%%%%%%%%%%%%%%%%%%%%%%%%%%%%%%%%%%%%%%%%%%%%%%%%%%%%%%%%%%%%%%%%%
We have reviewed and confirmed the basic results of \cite{Bernreuther:1994hq}
on $CP$ violation induced by non-standard Higgs exchange in the production of
$t\bar t$ pairs at the LHC, up to a few misprints. The results for the box
diagram have been further reduced to basic loop integrals.  Some modifications
of their observable $A_1=E_+-E_-$ (where $E_\pm$ are the lepton energies) have
been investigated. These modifications can in some kinematical situations
improve the signal-to-noise ratio, provided the $t\bar t$ invariant mass can
be determined with a reasonable precision.

For the observable $A_2={\bf p}_{\bar t}\cdot {\bf l}^+ 
-{\bf p}_{t}\cdot {\bf l}^-$, we find a much lower signal-to-noise ratio
than was given in \cite{Bernreuther:1994hq}. This observable thus appears
less suited for a search for $CP$ violation at the LHC (for details,
see sect.~\ref{Sec:observables-pp}).

In a minimal, $CP$-violating Two-Higgs-Doublet Model, where couplings and
masses are constrained and related by the Higgs potential and the Yukawa
couplings (``Model II''), the $CP$-violating effects can at the LHC be at the
per mille level, as given by Bernreuther and Brandenburg
\cite{Bernreuther:1994hq} (see also \cite{Schmidt:1992et}), provided that (i)
$\tan\beta$ is small ($\lsim 1$), and that (ii) there is {\it only one light
Higgs boson}.  If there are two neutral Higgs bosons below the $t\bar t$
threshold, the effects are severely reduced by cancellations among different
contributions.

Our study does not include QCD corrections.  Within pure QCD (no Higgs
exchange), next-to-leading-order corrections are known
\cite{Bernreuther:2001bx}.  They are not known for these $CP$-violating
effects.  Since we study ratios, one may hope that the dominant parts of the
QCD corrections cancel, but this remains to be seen.

Finally, we recall that there are other observables that we have not studied
(see, e.g.\ \cite{Bernreuther:1998qv,Brandenburg:1998xw}).  At the parton
level, the basic asymmetries are larger, of the order of $10^{-2}$, as
compared with the order $10^{-3}$ effects discussed here.  Thus, for some
other choice of observable, it is quite possible that a larger fraction of the
2HDM parameter space could be explored.

\goodbreak
\bigskip

%%%%%%%%%%%%%%%%%%%%%%%%%%%%%%%%%%%%%%%%%%%%%%%%%%%%%%%%%%%%%%%%%%%%%%%
{\bf Acknowledgments.}  We would like to thank P.N. Pandita and J. Sj\"olin
for collaboration in early stages of this work.  It is also a great pleasure
to thank W. Bernreuther and, in particular, A. Brandenburg, for very useful
discussions and correspondence. Also, we would like to thank I.F. Ginzburg and
M. Krawczyk for discussions on the 2HDM, and W. Hollik and M. Klasen for
advice on the ``LoopTools'' package.  This research has been supported by the
Research Council of Norway.
%%%%%%%%%%%%%%%%%%%%%%%%%%%%%%%%%%%%%%%%%%%%%%%%%%%%%%%%%%%%%%%%

%%%%%%%%%%%%%%%%%%%%%%%%%%%%%%%%%%%%%%%%%%%%%%%%%%%%%%%%%%%%%%%%%%%%%%%%%%%%%%
\section*{Appendix A}
\setcounter{equation}{0}
\renewcommand{\thesection}{A}
%%%%%%%%%%%%%%%%%%%%%%%%%%%%%%%%%%%%%%%%%%%%%%%%%%%%%%%%%%%%%%%%%%%%%%%%%%%%%%
In this appendix we collect complete results for the functions $b_{g1}^{CP}$,
$b_{g2}^{CP}$, $c_{g1}$, $c_{g2}$ of eq.~(\ref{eq:prod_mat}), correcting some
misprints in \cite{Bernreuther:1994hq} and also reducing completely the
contribution from the box diagram to four-, three-, and two-point scalar
integrals.  We define the following two-point functions:
\begin{align}
\label{J0011}
B_0(m_t^2,m_H^2,m_t^2) &=
\frac{1}{i\pi^2}
\int d^n \!q\frac{1}{
[(q-\half Q)^{2}-m_t^2][(q-\half P_t)^{2}-m_H^2]},
\\
\label{J1010}
B_0(\hat s,m_t^2,m_t^2) &=
\frac{1}{i\pi^2}
\int d^n \!q\frac{1}{[(q+\half Q)^{2}-m_t^2]
[(q-\half Q)^{2}-m_t^2]},
\\
\label{J0101}
B_0(\hat t,m_H^2,m_t^2) &=
\frac{1}{i\pi^2}
\int d^n \!q\frac{1}{[(q-\half P_g)^{2}-m_t^2]
[(q-\half P_t)^{2}-m_H^2]},
\end{align}
three-point functions:
\begin{align}
\label{J1110}
C_0(\hat{s},m_t^2,m_t^2,m_t^2) &=
\frac{1}{i\pi^2}
\int d^n \!q\frac{1}{[(q+\half Q)^{2}-m_t^2]
[(q-\half P_g)^{2}-m_t^2][(q-\half Q)^{2}-m_t^2]}, \\
\label{J1011}
C_0(\hat{s},m_H^2,m_t^2,m_t^2) &=
\frac{1}{i\pi^2}
\int d^n \!q\frac{1}{[(q+\half Q)^{2}-m_t^2]
[(q-\half Q)^{2}-m_t^2][(q-\half P_t)^{2}-m_H^2]}, \\
\label{J0111}
C_0(\hat{t},m_H^2,m_t^2,m_t^2) &=
\frac{1}{i\pi^2}
\int d^n \!q\frac{1}
{[(q-\half P_g)^{2}-m_t^2][(q-\half Q)^{2}-m_t^2][(q-\half P_t)^{2}-m_H^2]},
\end{align}
and the basic four-point function:
\begin{equation}
\label{J1111}
\!\!D_0(\hat t)=
\frac{1}{i\pi^2}
\int d^n \!q\frac{1}{[(q+\half Q)^{2}-m_t^2]
[(q-\half P_g)^{2}-m_t^2][(q-\half Q)^{2}-m_t^2][(q-\half P_t)^{2}-m_H^2]}.
\end{equation}
Here, $\hat s=Q^2$, $\hat t=-Q^2(1-\beta z)$.
Furthermore, for $\hat u=-Q^2(1+\beta z)$, the scalar
integrals $B_0(\hat u,m_H^2,m_t^2)$, $C_0(\hat{u},m_H^2,m_t^2,m_t^2)$
and $D_0(\hat u)$ are obtained from (\ref{J0101}), (\ref{J0111})
and (\ref{J1111}), respectively, with $P_g\to-P_g$ (or $P_t\to-P_t$).
Note that, in this notation of \cite{Bernreuther:1994hq}, the function
$C_0(\hat{s},m_t^2,m_t^2,m_t^2)$
can not be obtained from $C_0(\hat{s},m_H^2,m_t^2,m_t^2)$ by
a simple substitution.

We also define the abbreviations
\begin{align} 
D_F(\hat{t})&=m_H^2D_0(\hat{t})-(1-\beta z)C_0(\hat{t},m_H^2,m_t^2,m_t^2),
\nonumber\\
D_F(\hat{u})&=m_H^2D_0(\hat{u})-(1+\beta z)C_0(\hat{u},m_H^2,m_t^2,m_t^2),
\end{align}
where $\beta=(1-4m_t^2/\hat s)^{1/2}$ and $z=({\bf\hat P}_g\cdot{\bf\hat
P}_t)$, together with the UV-finite combination
\begin{align}
{\cal B}
&=9\beta z
[B_0(\hat{t},m_H^2,m_t^2)-B_0(\hat{u},m_H^2,m_t^2)]
+18z^2[B_0(\hat{s},m_t^2,m_t^2)-B_0(m_t^2,m_H^2,m_t^2)]\nonumber\\
&+7[B_0(\hat{t},m_H^2,m_t^2)+B_0(\hat{u},m_H^2,m_t^2)
-2B_0(m_t^2,m_H^2,m_t^2)],
\end{align}
and (see \cite{Bernreuther:1994hq})
\begin{align}
C_s(\hat{t})  \label{C_st} 
&=C_0(\hat{t},m_H^2,m_t^2,m_t^2)
+\frac{B_0(m_t^2,m_H^2,m_t^2)-B_0(\hat{t},m_H^2,m_t^2)}{m_t^2-\hat{t}}, \\
G(\hat{s}) \label{Eq:G_s} 
&=\frac{-[m_H^2C_0(\hat{s},m_H^2,m_t^2,m_t^2)
+B_0(\hat{s},m_t^2,m_t^2)-B_0(m_t^2,m_H^2,m_t^2)]}{\hat{s}\beta^2}.
\end{align}
Furthermore, $C_s(\hat{u})$ is obtained from eq.~(\ref{C_st})
by the replacement $\hat{t}\to\hat{u}$.
The factor
\begin{equation}
\kappa=\frac{m_t^2\sqrt{2}\;G_F\;g_s^4}{8\pi^2},
\end{equation}
where $g_s$ is the QCD coupling constant,
determines the over-all scale.

The functions $b_{g1}^{CP}$ and $c_{g1}$ are odd under 
$(z\to-z$, implying $\hat t\to \hat u$) whereas $b_{g2}^{CP}$ and $c_{g2}$
are even. In the following, the four- and three-point functions are
finite in four dimensions, as well as all combinations of two-point
functions.

For the box diagram (diagram ($c$) in fig.~\ref{Fig:3}), we find:
\begin{align} 
b^{(c)}_{g_1}&=\kappa\,\gamma_{CP}\,
              \frac{-m_t\,E_1}{192(1-\beta^2z^2)(1-z^2)\hat{s}\beta}
\nonumber\\
&\times\Bigl[
(7+9\beta z)\Bigl(
2z[\;\hat{s}\;(1-\beta z)^2+4m_H^2]\;\Im{D_F(\hat{t})}
\nonumber\\
&\hspace*{1.8cm}
+\{\hat{s}\beta z[\hat{s}\beta (1-\beta z)(1-2\beta z+z^2 )+4m_H^2(\beta -z)] 
\nonumber\\
&\hspace*{1.8cm}
- \hat{s}^2\beta(1-\beta z)(1-\beta^2)
\}\;\Im{D_0(\hat{t})}\Bigr) -(z\to -z;\; \hat t \to \hat u)\nonumber\\ 
%% C0(s,mH...
&+4\beta^2z\bigl[\hat{s}(7-18\beta^2z^2+7z^2)
-18m_H^2z^2(3-z^2)-9\hat{s}(1-\beta^2)
\bigr]\Im{C_0(\hat{s},m_H^2,m_t^2,m_t^2)}\nonumber\\
%%
%% C0(s,mt...
&+4z\bigl[\hat{s}(7-11\beta^2z^2)+28m_H^2
\bigr]\Im{C_0(\hat{s},m_t^2,m_t^2,m_t^2)}
%%
%% B0's...
-4\beta^2z(1-z^2)\Im{\cal B}
\Bigr], \\
\nonumber
b^{(c)}_{g_2}&=\kappa\,\gamma_{CP}\,
              \frac{m_t}{192(1-\beta^2z^2)(1-z^2)\hat{s}\beta}
\nonumber\\
&\times\Bigl[
(7+9\beta z)\bigl(
2[\;\hat{s}\;(1-\beta z)^2+4m_H^2][(1-z^2)m_t+E_1z^2]
\;\Im{D_F(\hat{t})}
\nonumber\\
&\hspace*{1.0cm}
+\{\hat{s}\beta[\hat{s}\beta(1-\beta z)(1-2\beta z+ z^2)+4m_H^2(\beta-z)]
[(1-z^2)m_t+E_1z^2] \nonumber\\
&\hspace*{1.0cm}
-\hat{s}^2E_1\beta z(1-\beta z)(1-\beta^2)\}\;\Im{D_0(\hat{t})}\bigr)
+(z\to -z;\; \hat t\to\hat u)\nonumber\\ 
%%
%% C0(s,mH...
&+4\beta^2\bigl\{[\hat{s}(7-18\beta^2z^2+7z^2)
-18m_H^2z^2(3-z^2)][(1-z^2)m_t+E_1z^2]\nonumber\\
&\hspace*{1cm}
-18z^2m_t[m_H^2(1-z^2)+2m_tE_1]
\bigr\}\Im{C_0(\hat{s},m_H^2,m_t^2,m_t^2)}\\
%%
%% C0(s,mt...
&+4[\hat{s}(7-11\beta^2z^2)+28m_H^2]
[(1-z^2)m_t+E_1z^2]\Im{C_0(\hat{s},m_t^2,m_t^2,m_t^2)}
\nonumber\\
%%
%% B0's...
&-4\beta^2(1-z^2)[(2-z^2)m_t+E_1z^2]\Im{\cal B}\nonumber
\Bigr],
\nonumber
\end{align}
\begin{align} 
c^{(c)}_{g_1}&=\kappa\,\gamma_{CP}\,
              \frac{m_t\,E_1}{192(1-\beta^2z^2)(1-z^2)\hat{s}\beta}
\nonumber\\
&\times\Bigl[
(7+9\beta z)\bigl\{
2z[\;\hat{s}\;(\beta^2z^2+2\beta^2-2\beta z-1)+4m_H^2]\;\Re{D_F(\hat{t})}
\nonumber\\
&\hspace*{1.8cm}
+\{\hat{s}\beta z[\hat{s}\beta(1-\beta z)(1-2\beta z+ z^2)+4m_H^2(\beta -z)]
\nonumber\\
&\hspace*{1.8cm} 
+ \hat{s}^2\beta(1-\beta z)(1-\beta^2)\}\;\Re{D_0(\hat{t})}\bigr\}
-(z\to-z;\; \hat t\to\hat u)
\\ 
%%
%%
%% C0(s,mH...
&+4\beta^2z\bigl[\hat{s}(7-18\beta^2z^2+7z^2)
-18m_H^2z^2(3-z^2)+9\hat{s}(1-\beta^2)
\bigr]\Re{C_0(\hat{s},m_H^2,m_t^2,m_t^2)}\nonumber\\
%%
%% C0(s,mt...
&-4z\bigl[\hat{s}(7+11\beta^2z^2)-28m_H^2-14\hat{s}\beta^2
\bigr]\Re{C_0(\hat{s},m_t^2,m_t^2,m_t^2)}
%%
%% B0's...
-4\beta^2z(1-z^2)\Re{\cal B}
\Bigr], \nonumber \\
%%%
c^{(c)}_{g_2}&\!=\kappa\,\gamma_{CP}\,
              \frac{-m_t}{192(1-\beta^2z^2)(1-z^2)\hat{s}\beta}
\nonumber\\
&\times\Bigl[
(7+9\beta z)\bigl(
2\{[\hat{s}(\beta^2z^2\!-\!2\beta z\!-\!1)+4m_H^2][(1\!-\!z^2)m_t+E_1z^2]
+2\hat{s}E_1\beta^2z^2\}\Re{D_F(\hat{t})}
\nonumber\\
&\hspace*{1.2cm}
+\{\hat{s}\beta[\hat{s}\beta(1-\beta z)(1-2\beta z+ z^2)+4m_H^2(\beta-z)]
[(1-z^2)m_t+E_1z^2] \nonumber\\
&\hspace*{1.2cm}
+\hat{s}^2E_1\beta z(1-\beta z)(1-\beta^2)\}\;\Re{D_0(\hat{t})}\bigr)
+(z\to-z;\; \hat t\to \hat u)\nonumber\\ 
%%
%% C0(s,mH...
&+4\beta^2\bigl\{[\hat{s}(7-18\beta^2z^2+7z^2)-18m_H^2z^2(3-z^2)]
[(1-z^2)m_t+E_1z^2]\nonumber\\
&\hspace*{1cm}
-18z^2m_t[m_H^2(1-z^2)-2m_tE_1]
\bigr\}\Re{C_0(\hat{s},m_H^2,m_t^2,m_t^2)}\\
%%
%% C0(s,mt...
&-4\{[\hat{s}(7+11\beta^2z^2)-28m_H^2]
[(1-z^2)m_t+E_1z^2]
-14\hat{s}E_1\beta^2z^2\}
\Re{C_0(\hat{s},m_t^2,m_t^2,m_t^2)}\nonumber\\
%%
%% B0's...
&-4\beta^2(1-z^2)[(2-z^2)m_t+E_1z^2]\Re{\cal B}.
\Bigr]\nonumber
\end{align}

%%%%%%%%%%%%%%%%%%%%%%%%%%%%%%%%%%%%%%%%%%%%%%%%%%%%%%%%%%%%%%%%%%%
From the vertex diagrams ($d$ and $e$ in fig.~\ref{Fig:3}):
\begin{align} 
c^{(d+e)}_{g_1}&\!=\kappa\,\gamma_{CP}\,
              \frac{-m_t\,E_1}{96(1-\beta^2z^2)}
\Bigl[
(7+9\beta z)\Bigl(2(1-\beta^2)\;C_0(\hat{t},m_H^2,m_t^2,m_t^2)
\nonumber\\
&\hspace*{2.4cm}+\frac{\beta z}{1-\beta z}\;(1+\beta^2z^2-2\beta^2)\;
C_s(\hat{t})\Bigr)-(z\to -z;\; \hat t \to \hat u)\Bigr],
\\
%%%%%%%%%%%% cg2 %%%%%%%
c^{(d+e)}_{g_2}&\!=\kappa\,\gamma_{CP}\,
              \frac{\beta\, m_t }{96(1-\beta^2z^2)}
\Bigl[
(7+9\beta z)\Bigl(\frac{2m_t}{E_1+m_t}(E_1+m_t-\beta zE_1)
C_0(\hat{t},m_H^2,m_t^2,m_t^2)\nonumber\\
&\hspace*{2.4cm}-\frac{m_t+E_1z^2-m_tz^2}{E_1^2(1-\beta z)}
(E_1^2-2m_t^2-\beta^2E_1^2z^2)C_s(\hat{t})\Bigr)
+(z\to -z;\; \hat t \to \hat u)\Bigr].
\nonumber
\end{align}

From the self-energy diagram (after fixing the signs):
\begin{align} 
c^{(f)}_{g_1}&\!=\kappa\,\gamma_{CP}\,
              \frac{-m_t}{192E_1(1-\beta^2z^2)}
\nonumber\\
&\times\Bigl[
(7+9\beta z)\frac{1-\beta^2}{1-\beta z}
\Bigl(B_0(m_t^2,m_H^2,m_t^2)-B_0(\hat{t},m_H^2,m_t^2)\Bigr)
-(z\to -z;\; \hat t \to \hat u)\Bigr],
\nonumber\\
%%%%%%%%%%%% cg2 %%%%%%%
c^{(f)}_{g_2}&\!=\kappa\,\gamma_{CP}\,
              \frac{\beta\,m_t}{192(1-\beta^2z^2)}
\Bigl[
(7+9\beta z)
\frac{m_t}{E_1+m_t}\frac{E_1+m_t-\beta zE_1}{E_1^2(1-\beta z)}
\nonumber\\
&\times\Bigl(B_0(m_t^2,m_H^2,m_t^2)-B_0(\hat{t},m_H^2,m_t^2)\Bigr)
+(z\to -z;\; \hat t \to \hat u)\Bigr].
\end{align}

From diagram ($g$):
\begin{align}  
b^{(g)}_{g_2}&\!=\kappa\,\gamma_{CP}\,
              \frac{-3m_t^2\beta^3z^2\Im{G(\hat{s})}}{8(1-\beta^2z^2)}
\nonumber\\
%%%%%%%%%%%% cg2 %%%%%%%
c^{(g)}_{g_2}&\!=\kappa\,\gamma_{CP}\,
             \frac{3m_t^2\beta^3z^2\Re{G(\hat{s})}}{8(1-\beta^2z^2)}.
\end{align}

Finally, from diagram ($h$):
\begin{align}  
b^{(h)}_{g_2}&\!=\kappa\,\gamma_{CP}\,
\frac{m_t^2}{(\hat{s}-m_H^2)^2+\Gamma_H^2m_H^2}\;
\frac{-\beta}{4(1-\beta^2z^2)} \\
&\times\Bigl[2m_t^2[\Gamma_H m_H\Re{C_0(\hat{s},m_t^2,m_t^2,m_t^2)}
-(\hat{s}-m_H^2)\Im{C_0(\hat{s},m_t^2,m_t^2,m_t^2)}]+\Gamma_H m_H
\Bigr],
\nonumber\\
%%%%%%%%%%%% cg2 %%%%%%%
c^{(h)}_{g_2}&\!=\kappa\,\gamma_{CP}\,
\frac{m_t^2}{(\hat{s}-m_H^2)^2+\Gamma_H^2m_H^2}
\Bigl[
\frac{\beta}{16(1-\beta^2z^2)}\,\Bigl(2\hat{s}(1+\beta^2)\\
&\quad\times
[(\hat{s}-m_H^2)\Re{C_0(\hat{s},m_t^2,m_t^2,m_t^2)}
+\Gamma_H m_H\Im{C_0(\hat{s},m_t^2,m_t^2,m_t^2)}]
-4(\hat{s}-m_H^2)\Bigr)\nonumber\\
&\quad+\frac{3}{32}\frac{m_t^2\sqrt{2}\;G_F}{8\pi^2}
[2\hat{s}^3\beta \tilde a^2|C_0(\hat{s},m_t^2,m_t^2,m_t^2)|^2
+2\hat{s}\beta a^2
|2-\hat{s}\beta^2C_0(\hat{s},m_t^2,m_t^2,m_t^2)|^2]\Bigr], \nonumber
\end{align}
where $\Gamma_H$ is the width of the Higgs boson calculated by summing
the partial widths for $H\to W^+W^-, ZZ, t\bar t$ in the version of
2HDM we considered. 
\bigskip\goodbreak

%%%%%%%%%%%%%%%%%%%%%%%%%%%%%%%%%%%%%%%%%%%%%%%%%%%%%%%%%%%%%%%%%%%%%%%%

\end{document}